\documentclass[12pt,preprint]{aastex}

\newcommand{\kev}{keV}
\newcommand{\fe}{Fe~K$\alpha$}
\newcommand{\etal}{et al.}
\newcommand{\nustar}{\textit{NuSTAR}}

\newcommand{\chandra}{\textit{Chandra}}
\newcommand{\xmm}{\textit{XMM-Newton}}

\shorttitle{Hard X-ray AGN Number Counts}
\shortauthors{Ballantyne \etal}

\begin{document} 

\title{Lifting the Veil on Obscured Accretion: Active Galactic Nuclei
  Number Counts and Survey Strategies for Imaging Hard X-ray Missions}


\author{D. R. Ballantyne\altaffilmark{1},
  A. R. Draper\altaffilmark{1}, K. K. Madsen\altaffilmark{2},
  J. R. Rigby\altaffilmark{3,4,5}, and E. Treister\altaffilmark{6,7,8}}
\altaffiltext{1}{Center for Relativistic Astrophysics, School of Physics,
  Georgia Institute of Technology, Atlanta, GA 30332;
  david.ballantyne@physics.gatech.edu}
\altaffiltext{2}{Space Radiation Laboratory, California Institute of
  Technology, MC: 220-47, Pasadena, CA 91125}
\altaffiltext{3}{Carnegie Observatories, Carnegie Institution of
  Washington, 813 Santa Barbara St., Pasadena, CA 91101}
\altaffiltext{4}{Carnegie Fellow}
\altaffiltext{5}{Current Address: NASA Goddard Space Flight Center,
  Code 665, Greenbelt MD 20771}
\altaffiltext{6}{Institute for Astronomy, 2680 Woodlawn Drive,
  University of Hawaii, Honolulu, HI 96822}
\altaffiltext{7}{Universidad de Concepcion, Departamento de
  Astronomia, Casilla 160-C, Concepcion, Chile}
\altaffiltext{8}{\chandra/Einstein Fellow}

\begin{abstract}
Finding and characterizing the population of active galactic nuclei
(AGNs) that produces the X-ray background (XRB) is necessary to
connect the history of accretion to observations of galaxy evolution
at longer wavelengths. The year 2012 will see the deployment of the
first hard X-ray imaging telescope that, through deep extragalactic
surveys, will be able to measure the AGN population at the energies
where the XRB peaks ($\sim 20$--30~keV). Here, we present predictions
of AGN number counts in three hard X-ray bandpasses: $6$--$10$~keV,
$10$--$30$~keV and $30$--$60$~keV. Separate predictions are presented
for the number counts of Compton thick AGNs, the most heavily obscured
active galaxies. The number counts are calculated for five different
models of the XRB that differ in the assumed hard X-ray luminosity
function, the evolution of the Compton thick AGNs, and the underlying
AGN spectral model. The majority of the hard X-ray number counts will
be Compton thin AGNs, but there is a $>10\times$ increase in the
Compton thick number counts from the $6$--$10$~keV to the $10$--$30$~keV band. The Compton thick population
show enough variation that a hard X-ray number counts
measurement will constrain the models. The computed number
counts are used to consider various survey strategies for the \nustar\
mission, assuming a total exposure time of $6.2$~Ms. We find that
multiple surveys will allow a measurement of Compton thick
evolution. The predictions presented here should be useful for all
future imaging hard X-ray missions.
\end{abstract}


\section{Introduction}
\label{sect:intro}
The vast majority of the growth of supermassive black holes (SMBHs) is
driven by accretion \citep[e.g.,][]{marc04,mh08,dim08}. Therefore, a complete census of accreting
SMBHs throughout cosmic time is necessary to quantify the efficiency
of accretion \citep[e.g.,][]{solt82,yt02,shak04,yl08}, which, in turn, elucidates the connection
between black hole growth and galaxy evolution
\citep[e.g.,][]{hco04,hjb07,hh09,sbh09}. All efficiently
accreting SMBHs are intrinsically luminous X-ray sources, while the underlying host
galaxy is, in general, a much fainter X-ray emitter. The cosmic X-ray
Background (XRB) is then naturally interpreted as being the integral
emission of all accreting SMBHs in the universe, and the 
hard slope of the XRB spectrum indicates that most of the AGNs are obscured
behind substantial columns of gas and dust in the host galaxy
\citep[e.g.,][]{sw89,com95,ueda03,gill01,gch07}. This expectation was spectacularly confirmed by the numerous
deep surveys performed by \chandra\ and \xmm\ in the $2$--$10$~\kev\
band that resolved $>80$\% of the XRB into individual objects \citep[e.g.,][]{bh05,wor05}. Multiwavelength follow-up observations of these sources
show that a large fraction of them are obscured, type 2, AGNs at
redshifts of $z \sim 1$ \citep{alex01,toz01,bar02,stern02,bar05,brusa10}.

However, summing the emission from all the AGNs detected by \chandra\
and other missions working below 10~\kev\ cannot account for the
observed peak of the XRB at $\sim 30$~\kev\ \citep[e.g.,][]{gch07}. The missing AGNs
are likely dominated by Compton thick sources, those behind obscuring
column densities $N_{\mathrm{H}} \gtrsim \sigma_{\mathrm{T}}^{-1} \approx
10^{24}$~cm$^{-2}$. The combination of photoelectric absorption and
Compton scattering reduces the observed flux of these AGNs to such a degree that, if they lie at
any reasonable cosmological distance, they are invisible even in the
deepest \chandra\ or \xmm\ surveys \citep{bh05}. There are currently three strategies
that are being employed to discover and identify Compton thick AGNs. The first is
to look for the `waste heat' being re-radiated by the absorber in the
mid-infrared. Several groups have uncovered a population of Compton
thick candidates at $z \gtrsim 1$ using this idea \citep{stern05,ah06,pol06,daddi07,alex08,fiore08,fiore09}, but the constraints
are complicated by the necessity to separate out contributions from
obscured star formation \citep[e.g.,][]{mush04,don08,eck10,geo10}. The second method to detect Compton
thick AGNs is to search at hard X-ray energies ($\gtrsim 20$~\kev)
where the sources are less affected by obscuration (although still suffer from Compton scattering
losses). The hard X-ray detectors on board \textit{BeppoSAX},
\textit{Swift}, and \textit{INTEGRAL} have been able to reveal local
(i.e., $z \approx 0$) Compton thick AGNs through this method
\citep[e.g.,][]{bass99,vc02,mal09a}. Recent
surveys by the latter two missions have measured the fraction of all
AGNs that are Compton thick at $z \approx 0$ to be $\sim10$--$20$\%, depending on the
luminosity range considered and the details of the source selection
\citep{saz07,tuell08,mal09b,tuv09,winter09,bur11}. Finally, Compton thick AGNs may
also be identified in pointed X-ray observations by searching for
objects with specific optical characteristics \citep{gilli11}, or those that have 
very large \fe\ equivalent widths
\citep[e.g.,][]{leven06,lamassa09,com11,feru11}. However, these
observations require very long exposure times and the translation from
\fe\ line measurement to column density depends on
model-dependent details such as the geometry and metallicity of the absorber
\citep[e.g.][]{my09}.

The hard X-ray instruments flown in previous missions have had fairly
poor sensitivity when compared to detectors at lower energies. Therefore, there is very
little known about the Compton thick AGN population at $z >
0$. Intriguingly, if one takes the infrared estimates of Compton thick
densities at high redshift at face value, then there must be strong
evolution of the Compton thick population \citep{db10,tre10}. This
evolution might indicate that `Compton thickness' is an unique phase of
AGN evolution and is tied to specific Eddington ratios or merger
events \citep{db10,tre10}. Therefore, discovering the missing Compton
thick AGNs beyond the local Universe and tracing their evolution is vital to
understanding a host of problems related to black hole fueling and
galaxy evolution.  

In 2012 the \textit{Nuclear Spectroscopic Telescope
Array}\footnote{\texttt{http://www.nustar.caltech.edu/}} (\nustar; \citealt{harr10}) will be launched as part of NASA's Small Explorer
Program. This mission will be the first ever focusing hard X-ray
telescope covering an energy range of $6$--$80$~\kev. It will
therefore be imaging the sky at the energies where the XRB peaks and
the Compton thick sources are most visible. One of the primary
objectives of the baseline two-year \nustar\ mission is to perform extragalactic surveys to describe
the AGN population and XRB emission at these energies. Other hard
X-ray imaging missions, such
as \textit{Astro-H} (Takahashi \etal\ 2010; launching in 2013/2014)
and the proposed \textit{New Hard X-ray Mission} \citep{tag10}, will
follow later this decade. The data produced by these missions will
provide crucial tests of models of AGN and galaxy
evolution.

Thus, in light of the upcoming launch of \nustar, there is a need for
models of AGN evolution, and, specifically, Compton thick evolution,
that can be used to compare against future hard X-ray data. In
addition, model predictions can also be used to assist the planning of
the hard X-ray extragalactic surveys to ensure that a
reasonable number of Compton thick AGNs are detected at several
different redshifts. To be the most effective, several models,
spanning a range of parameters, should be available to compare against
the survey data with the goal that many of them will be eliminated. As
a first step, this paper predicts the hard X-ray number counts for
three different measurements of the AGN hard X-ray luminosity
function with each requiring a different Compton
thick fraction. In addition, we also predict counts for a
model where the Compton thick fraction is a function of AGN Eddington
ratio, and therefore varies with redshift and AGN luminosity. Finally,
we also present a conservative model that predicts the smallest
fraction of Compton thick AGNs consistent with the available data. Our results are specifically designed
for comparison with the upcoming hard X-ray surveys, and are the first
to show the effects of the different model parameters on the AGN and
Compton thick counts.

The next section describes the
various XRB models used to calculate the Compton thick and total AGN
number counts. The counts are presented in
Section~\ref{sect:res}. Sect.~\ref{sect:surveys} then applies these
results to the design of deep hard X-ray surveys, with a specific
example for \nustar. Our conclusions for the
\nustar\ mission and for other future hard X-ray missions are summarized in Section~\ref{sect:concl}. This
paper assumes the following cosmological parameters:
$H_0=70$~km~s$^{-1}$~Mpc$^{-1}$, $\Omega_{\Lambda}=0.7$, and
$\Omega_{m}=0.3$.

\section{Calculation of Predicted Number Counts}
\label{sect:numcounts}
The integrated AGN number counts, $N(>S)$, in a specific
energy band are 
\begin{equation}
N(>S) = \frac {c}{H_0}\int^{z_{\mathrm{max}}}_{z_{\mathrm{min}}} \int^{\log
  L_X\mathrm{(max)}}_{\mathrm{max(}\log L_X\mathrm{(min)},\log L_X\mathrm{(S_{N_{\mathrm{H}}}))}}
\frac{d\Phi(L_X,z)}{d\log L_X}\frac{d_l^2}{(1+z)^2 (\Omega_m
  (1+z)^3 + \Omega_{\Lambda})^{1/2}}d\log L_X dz,
\label{eq:NofS}
\end{equation}
where $d\Phi (L_X,z)/d\log L_X$ is a given hard X-ray luminosity
function (HXLF), $d_l$ is the luminosity distance to redshift $z$, and $L_X(S_{N_{\mathrm{H}}})$ is
the unabsorbed rest-frame $2$--$10$~\kev\ luminosity that gives an observed frame
flux $S_{N_{\mathrm{H}}}$ for a source at $z$ averaged over the
$N_{\mathrm{H}}$ distribution for that $L_{X}$ and $z$. The integrals are evaluated from
$z_{\mathrm{min}}=0$ to $z_{\mathrm{max}}=5$, and $\log (L_X\mathrm{(min)}/\mathrm{erg\ s}^{-1})=41.5$ to $\log (L_X\mathrm{(max)}/\mathrm{erg\ s}^{-1})=48$. The integrated counts
are computed in three energy bands\footnote{To assist comparisons with
  previous observations and models, Appendix~A shows predictions in the
  $5$--$10$~\kev\ band.}: 6--10 keV, 10--30 keV and 30--60
keV. As described below, the calculation of the expected number counts
relies on a number of quantities which are poorly constrained or
subject to significant uncertainties.

\subsection{Unabsorbed AGN Spectrum}
\label{sub:agnspect}
The conversion between luminosity and flux requires a spectral model
for the average AGN population at a given $L_X$ and $z$. The assumed
unabsorbed rest-frame AGN spectrum is comprised of a power-law with photon
index $\Gamma$ that varies from source to source \citep[e.g.,][]{winter09}, an exponential cutoff at $E_{\rm cut}=250$~keV, and a
reflection component. The distribution of the power-law cutoff energies is not
observationally constrained and will be an important measurement for
the upcoming hard X-ray missions. However, \citet{gch07} find that
including a range of cut-off energies in their X-ray background
synthesis model has a negligible effect on the results; therefore, we
do not consider this parameter to be a significant source of
uncertainty in the predictions.

The reflection component hardens the spectrum above
$\ga 10$~\kev\ and produces an \fe\ line. A reflection spectrum may
arise from either the accretion disk or from cold, dense gas at large
distances from the black hole \citep[e.g.,][]{rf05,my09}. Thus, the strength of the
reflection component in an AGN spectrum may vary significantly
from source to source, although it is known that, on average, it
cannot be very large \citep{gan07,bal10}. In addition, there is observational
evidence that the strength of the reflection spectrum from the distant
reprocessor is anti-correlated with the
luminosity \citep{bia07,shu10}. 

\subsection{$N_{\mathrm{H}}$ Distribution and Compton Thick AGNs}
\label{sub:nhdist}
The basic spectral model will suffer some obscuration that depends on
the fraction of obscured AGNs at a specific $L_X$ and $z$, as well as
the distribution of obscuring column densities. The fraction of
absorbed, or type 2, AGNs ($f_2$) is defined as those objects observed
through column densities $N_{\mathrm{H}} \geq 10^{22}$~cm$^{-2}$. This
fraction is observed to decrease with AGN luminosity \citep[e.g.,][]{ueda03,bar05,laf05,tuell08}, and,
tentatively, increase with $z$ \citep{laf05,bem06,tu06,has08}. The distribution of column
densities that comprise the obscured and unobscured population has
been measured in the local universe, or in a redshift and luminosity integrated sense
\citep{com95,risaliti99,ueda03,dwelly05,tozzi06}. However, it is possible (perhaps likely) that the
distribution of column densities will also depend on $L_X$ and
$z$. 

Previous deep X-ray surveys by \chandra\ and \xmm\ are not
sensitive to most Compton thick AGNs; therefore the space density and
evolution of these objects represents the largest uncertainty in the
census of AGN. Additional unobserved objects have to be added
to the HXLF in order to boost the integrated emission to fit the XRB
spectrum, and/or the measured $z=0$ Compton thick space
density \citep{db09}. As mentioned above, Compton thick AGNs are obscured by
columns $N_{\mathrm{H}} \gtrsim 10^{24}$~cm$^{-2}$, but once
$N_{\mathrm{H}}$ reaches a few times $10^{24}$~cm$^{-2}$ the
transmitted continuum becomes severely suppressed
\citep{matt99}. Indeed, extreme Compton thick AGNs with
$N_{\mathrm{H}} \gtrsim 10^{25}$~cm$^{-2}$ will be, for all practical
purposes, invisible even in the deepest hard X-ray
surveys. Consequently, such objects cannot provide a significant
contribution to the XRB, and this paper defines Compton thick
AGNs as those with columns $24 \leq \log N_{\mathrm{H}} \leq 25$. 

The simplest method for including Compton thick AGNs into a XRB model is as a straightforward extension
of the Compton thin population that evolve with other obscured AGNs
(i.e., along with $f_2$). This is quantified by the Compton thick
fraction, $f_{\mathrm{CT}}$, defined as the fraction of all type 2
AGNs that are Compton thick. For example, $f_{\mathrm{CT}} = 0.5$
means that there are as many Compton thick AGNs as Compton thin type 2s.

\subsection{X-ray Luminosity Functions}
\label{sub:lfs}
In recent years there have been several measurements of the
HXLF that have significantly different predictions for the value and redshift evolution
of the faint end slope of the HXLF
\citep{ueda03,bar05,laf05,silv08,eb09,yen09,aird10}. For a given
spectral model and $N_{\mathrm{H}}$ distribution, these differences in
the HXLF can result in factors of 2--3 variations in the number density of
Compton thick AGNs \citep{db09}.

\subsection{Model Definitions}
\label{sub:models}
Given all the uncertainties associated with the calculation of number
   counts, we consider five separate models to determine the impact of our
   ignorance of many of the above parameters. The most interesting
   physical parameters that will likely be constrained by the hard X-ray
   number counts are the HXLF and Compton thick fraction.  Therefore,
   we consider three XRB models with a fixed Compton thick fraction that
   differ only in the assumed HXLF. The fourth XRB model uses one of
   the test HXLFs (specifically, the \citet{ueda03} HXLF), but assumes a more complicated
   form for the Compton thick evolution, where the Compton thick fraction
   depends on the Eddington ratio. The fifth and final model provides
   a lower-limit to the number of Compton thick AGNs. This model uses the
   \citet{ueda03} HXLF and a fixed Compton thick fraction, but assumes a
   brighter AGN spectrum and the minimum allowable Compton thick
   density. Thus, this model predicts the most conservative view of
   the Compton thick population that is consistent with the available
   data\footnote{Interested readers can perform a similar
   test by computing the number counts for the \citet{gch07} XRB
   synthesis model using their webtool at
   \texttt{http://www.bo.astro.it/$\sim$gilli/counts.html}. Note that the
   Compton thick fraction of this model is similar to the HXLF
   variable models used here. As Appendix~A shows, the \citet{gch07} predictions are
   within the range of those studied in this paper, and do not change
   any of our conclusions.}.

\subsubsection{HXLF Variable Models}
\label{subsub:hxlfmods}
In the following models, the reflection spectrum is calculated
using the `reflion' model \citep{rf05} assuming a solar Fe
abundance. This reflection spectrum is averaged over all viewing
angles, and is
added to the power law such that the equivalent width of the narrow \fe\ line agrees with the observed X-ray Baldwin effect
\citep{bia07}. The corresponding reflection fraction drops from $\sim
1$ at low luminosities to close to zero at high luminosities. To account for the observed distribution of photon
indices, both the power-law and the reflection spectrum are Gaussian
averaged about $\Gamma = 1.9$ with $\Gamma_{\rm min} = 1.5$, 
$\Gamma_{\rm max} = 2.3$ and $\sigma_{\Gamma} = 0.2$ \citep[e.g.,][]{gch07}.

The fraction of obscured AGNs, $f_2$, is assumed to be dependent on both
redshift and $L_X$, such that $f_2 \propto (1+z)^a(\log L_X)^{-b}$
where $a=0.4$\footnote{If $a=0.62$ and
  continues out to $z=2$ \citep{has08}, then the $10$--$30$~\kev\
  Compton thick counts are increased
  by $\la 13$\% at $1 < z < 2$, but are reduced by $\la 9$\%
  at $z < 0.5$.} \citep{bem06,tu06} and $b=4.7$. The normalization of
$f_2$ is set by fixing the type 2 to type 1 ratio at $4:1$ for $z=0$
and $\log (L_X/\mathrm{erg\ s}^{-1}) =41.5$, as determined from 
observations \citep{mr95} and XRB modeling \citep{bem06,gch07}. The redshift
evolution is halted at $z=1$ corresponding to the flattening evolution
of the cosmic star formation rate density \citep[e.g.,][]{hb06}. The unabsorbed type 1 sources
are evenly distributed over the columns $\log N_H = 20, 20.5, 21$, and
$21.5$, while Compton thin type 2 sources are evenly distributed over the
columns $\log N_H = 22, 22.5, 23$, and $23.5$.

The Compton thick fraction, $f_{\mathrm{CT}}$, is determined by requiring
the Compton thick AGN space density at $z \approx 0$ with $\log
(L_X/\mathrm{erg\ s}^{-1})
> 43$ to be $\approx 3.6\times 10^{-6}$~Mpc$^{-3}$, in agreement with the
absorption corrected local space density measurement from \textit{Swift}/BAT and
\textit{INTEGRAL} \citep{tuv09,rig09,db10}. This value is very similar
to the density predicted by the \citet{gch07} model
\citep[e.g.,][]{alex08}, which, judging from Fig.~3 in the
paper by \citet{tuv09}, suggests that
$\sim 15$\% of $z \sim 0$ AGNs in this luminosity range are
Compton thick. The Compton thick
AGNs are evenly distributed over the columns $\log N_H = 24, 24.5$,
and $25$, consistent with the limits from local observations \citep{tuv09}, and the Compton thick absorption profiles are taken from the
Monte-Carlo models of \citet{matt99}. A scattered reflection spectrum
with a luminosity of $2$\% of the intrinsic luminosity is used in place of the transmitted absorption spectrum for the $\log
N_H =25$ spectral model, and is added to the $\log N_H =24.5$ spectrum
\citep[e.g.,][]{gch07}. 

The three HXLFs used in these models (in order of increasing Compton thick
fraction) are the ones published by Ueda \etal\ (2003; $f_{\mathrm{CT}}=0.3$,
  blue lines in all figures), La Franca \etal\ (2005; $f_{\mathrm{CT}}=0.4$,
  green lines) and Aird \etal\ (2010; $f_{\mathrm{CT}}=0.5$, red lines). The \citet{ueda03} and
\citet{laf05} HXLFs are best modeled by luminosity dependent density
evolution (LDDE), while \citet{aird10} find that luminosity and density
evolution (LADE) is the best description of their data.

\subsubsection{Compton Thick Evolution Model}
\label{subsub:compmod}
 In addition to the above three models, we also consider
the scenario where the Compton thick fraction ($f_{\mathrm{CT}}$) evolves separately from the less
obscured AGN. Specifically, we make use of the composite model of
\citet{db10} who assumed that $f_{\mathrm{CT}}$ was a function of the AGN
Eddington ratio. After considering several scenarios and comparing to
the high-$z$, IR-derived Compton thick space densities, \citet{db10}
estimated that Compton thick AGNs make up $\sim 86$\% of all AGNs with Eddington ratios
greater than 0.9, $\sim 60$\% of AGNs with Eddington ratios less than
0.01, and $0$\% of AGNs at intermediate Eddington ratios. This model makes use of the
\citet{ueda03} HXLF and recovers the observed $z=0$ Compton thick
density \citep{db10}. All other parameters are identical to the previously
described models. This model is shown as the black line in all
figures.

\subsubsection{The Conservative Compton Thick Model}
\label{subsub:altmod}
We define a model, described by \citet{tuv09}, that provides a lower-limit to the Compton thick population that is consistent with all
available data. This model
assumes a constant power-law with $\Gamma=1.9$ and a cutoff energy of $300$~\kev, attenuated by photoelectric
absorption and a constant Compton reflection component as given by
\citet{magdziarz95}. The reflecting material is assumed to provide a reflection fraction of 1.0, has twice the solar iron
abundance, and has an average inclination angle to the line of sight
of 45~deg. The larger iron abundance and constant reflection fraction
result in a slighter more luminous AGN spectrum above 10~\kev\ when
compared to the models described in Sect.~\ref{subsub:hxlfmods}
and~\ref{subsub:compmod}. 

The $N_{\mathrm{H}}$ distribution employed here is the one shown in
figure 2 of \citet{tuv09} and was originally
  derived from a torus geometry corresponding to an AGN unification
  model with
  an aspect ratio matching the local ratio of Type 2 to Type 1 Seyfert
  galaxies \citep{treister04}. However, it is fully consistent with
  the observed distributions in various hard X-ray surveys
  (\citealp{com95,risaliti99,ueda03,dwelly05,tozzi06}). In this
  $N_{\mathrm{H}}$ distribution the majority of the Compton thick AGNs
  have $N_{\mathrm{H}} < 10^{24.5}$~cm$^{-2}$, which results in a
  more luminous average Compton thick spectrum than used in the models
  described above. Since this paper is focused on either the Compton
  thick AGNs, or on the total AGN population, the remaining
  differences in the two $N_{\mathrm{H}}$ distributions have no impact
  on the results. The luminosity
dependence of the obscured AGN fraction $f_2$ falls linearly from
$f_2=1$ to $0$ between $L_X=10^{42}$ and
$3\times 10^{46}$~erg~s$^{-1}$ \citep{treister05}. The redshift
dependence of $f_2$ is the same as the other models, and the
\citet{ueda03} HXLF is assumed.

The Compton thick fraction of this model is set to be consistent with
the observed fraction of AGNs at $z \approx 0$ that are Compton thick, 
as deduced from hard X-ray surveys by \textit{Swift} and
\textit{INTEGRAL} \citep{saz07,tuell08,tuv09}. This corresponds to a
fraction of all $z\sim 0$ AGNs that are Compton thick of $\sim 10$\%
(\citealt{tuv09}; Fig. 3), and is likely to
be a strict lower-limit due to the observational biases of those
surveys. The extinction corrected density used to normalize the
Compton thick fraction in the models described in
Sect.~\ref{subsub:hxlfmods} attempts to partially address those
biases \citep{rig09}. Thus, this model presents the most
conservative view of the Compton thick population, and is shown as the
cyan line in all plots. 

\subsection{Summary of Models}
\begin{figure}
\includegraphics[angle=-90,width=0.95\textwidth]{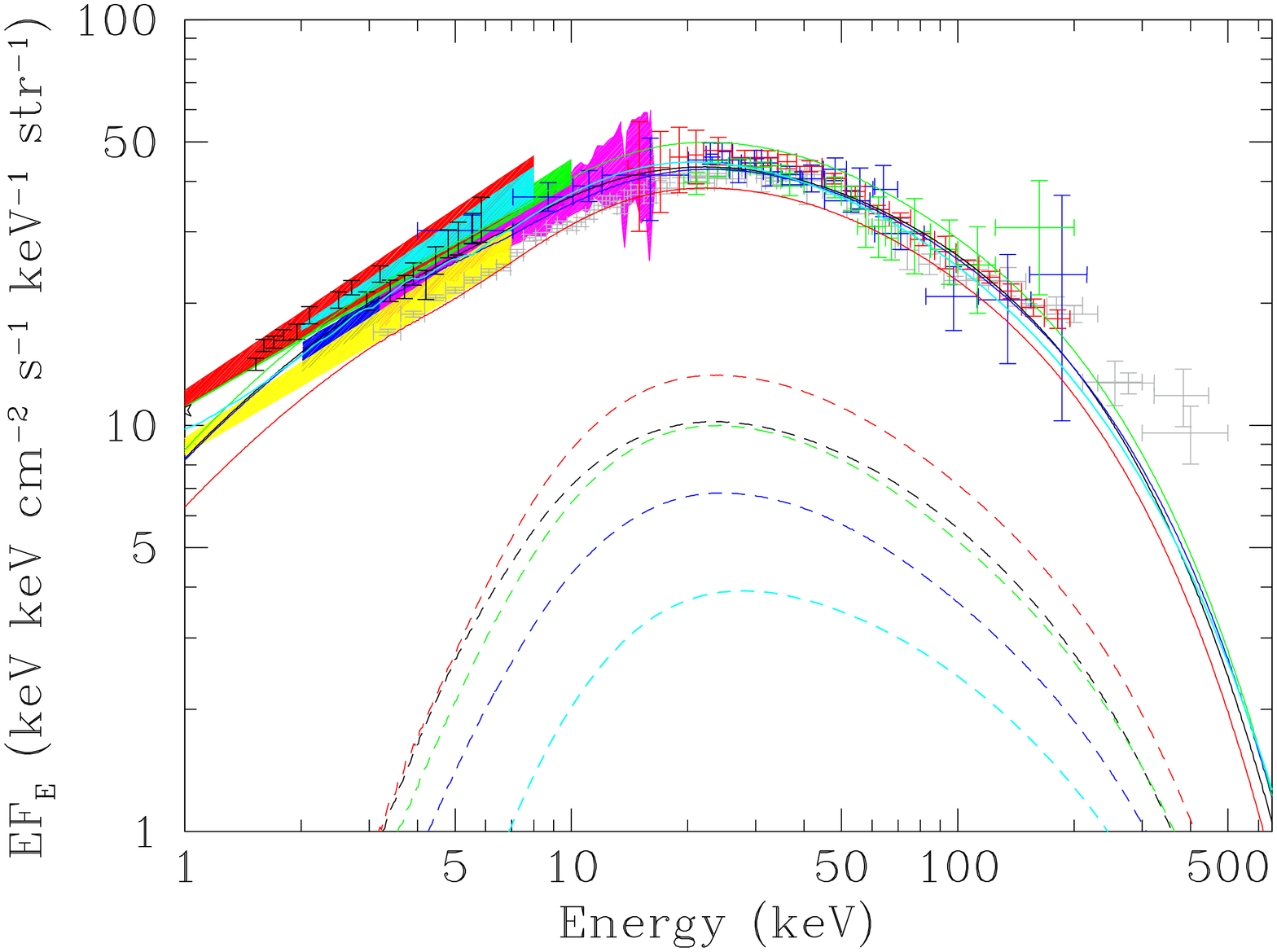}
\caption{The solid lines show the X-ray background spectra predicted
  by the five AGN evolution models considered here: \citet{ueda03}
  HXLF (blue), \citet{laf05} HXLF (green), \citet{aird10} HXLF (red),
  the \citet{db10} composite model (black), and the conservative Compton thick
  model \citep{tuv09} (cyan). The dashed lines plot the
Compton thick contribution to the X-ray background. The X-ray
background data are from the following instruments: blue: {\em ASCA}
GIS (Kushino \etal\ 2002); magenta: {\em Rossi X-ray Timing Explorer}
({\em RXTE}; Revnivtsev \etal\ 2003); green: XMM-{\em Newton} (Lumb
\etal\ 2002); red: {\em BeppoSAX} (Vecchi \etal\ 1999); yellow: {\em
ASCA SIS} (Gendreau \etal\ 1995); cyan: XMM-{\em Newton} (De Luca \&
Molendi 2004); grey data: {\em HEAO}-1 (Gruber \etal\ 1999); blue
data: {\em INTEGRAL} (Churazov \etal\ 2007); red data: {\em Swift}/BAT
(Ajello \etal\ 2008); black data: {\em Swift}/XRT (Moretti \etal\
2009); green data: {\em INTEGRAL} (T\"urler \etal\ 2010).}
\label{fig:cxrb}
\end{figure}

\begin{figure}
\includegraphics[angle=-90,width=0.95\textwidth]{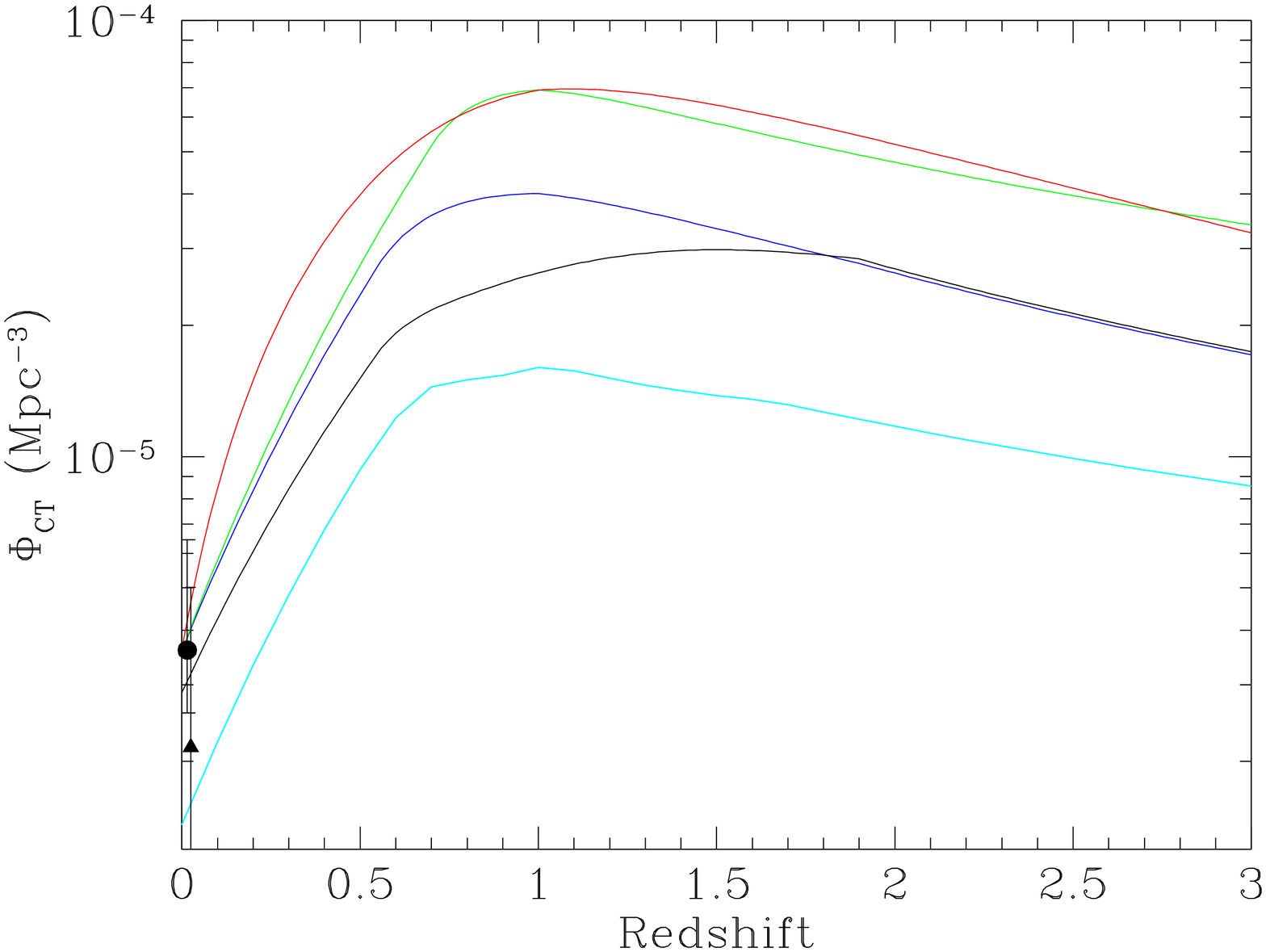}
\caption{Predicted space density of Compton thick AGNs with $\log
  (L_X/\mathrm{erg\ s}^{-1}) > 43$ as a function of $z$. The different models are distinguished by
the line color: \citet{ueda03} HXLF (blue), \citet{laf05} HXLF
(green), \citet{aird10} HXLF (red),
  the \citet{db10} composite model (black), and the conservative
  Compton thick model \citep{tuv09} (cyan). The triangle denotes the local Compton thick density
  calculated by \citet{tuv09}, while the circle shows the same density
  after correcting for the flux-luminosity relation reported by \citet{rig09}.}
\label{fig:ctevol}
\end{figure}

Predictions of the hard X-ray number counts rely on several parameters
or distributions that are only partially observationally
constrained. Thus, five separate models were defined to fully explore
the sensitivity of the results to these
assumptions. Of the five, three explored differences in the measured
AGN HXLF (Sect.~\ref{subsub:hxlfmods}), one considered a model where
the Compton thick fraction was a function of AGN accretion rate
(Sect.~\ref{subsub:compmod}), and one provides the most conservative
view of the
Compton thick population (Sect.~\ref{subsub:altmod}). Figure~\ref{fig:cxrb} shows the X-ray background
spectrum, as well as the contribution from Compton thick AGNs for these
five models. All models provide an adequate fit to the
X-ray background spectrum, which nicely illustrates the degeneracy of
synthesis modeling. The redshift evolution of the $\log
(L_X/\mathrm{erg\ s}^{-1}) > 43$ Compton thick space density predicted by the models is shown in
figure~\ref{fig:ctevol}. The differences in the three different HXLFs
are easily observed with the Compton thick density increasing with the
necessary $f_{\mathrm{CT}}$. At $z \la 1$, the \citet{ueda03} HXLF requires
the least amount of Compton thick AGNs, and the \citet{aird10} HXLF
has the most. Above $z \ga 1$, the different evolutions of the HXLFs
become apparent and the \citet{laf05} and \citet{aird10} HXLFs predict
a similar Compton thick density. The brighter spectral model and lower
Compton thick fraction employed
in the \citet{tuv09} model results in a low Compton thick density at
all $z$. Finally, the evolving Compton thick model predicts a
small fraction of high luminosity Compton thick AGNs at $z < 1$
because there are very few high Eddington ratio AGNs to host
such obscured objects. Thus, there is a significant difference in the
predicted densities of Compton thick objects at $z < 1$, implying that
deep hard X-ray surveys may be capable to distinguish among these
models.

\section{Results}
\label{sect:res}
\subsection{Number Counts}
\label{sub:counts}
Figures~\ref{fig:6to10counts},~\ref{fig:10to30counts}
and~\ref{fig:30to60counts} show the predicted $N(>S)$ distributions
for multiple redshift ranges in the $6$--$10$~\kev, $10$--$30$~\kev\
and $30$--$60$~\kev\ bands, respectively. The solid lines plot the
counts for all AGN, while the dashed lines present the predictions for
Compton thick sources. The colors differentiate between the various
HXLFs and Compton thick evolutions: \citet{laf05} HXLF (green), \citet{aird10} HXLF (red), the
composite model of \citet{db10} (black), and the \citet{tuv09} model
(cyan). For clarity, the model based on the \citet{ueda03} HXLF has
been omitted from the plots. Recall that both the
\citet{db10} and the \citet{tuv09} make use of the \citet{ueda03}
HXLF. 

\subsubsection{All AGNs}
\label{subsub:allagns}
Concentrating initially on the counts of all AGNs (solid lines), we
see that the numbers are very similar to those observed in the softer
energy bands by \chandra\ and \xmm\
\citep[e.g.,][]{ros02,leh05,mat08}. This is not surprising: most AGNs
are only moderately obscured and so would appear in both
$2$--$8$~\kev\ and $10$--$30$~\kev\ surveys. These results
imply that most of the sources found in hard X-ray surveys will have
\chandra\ and/or \xmm\ counterparts with similar fluxes, especially
for AGNs at higher $z$ where the K-correction has moved the absorption
out of the \chandra\ or \xmm\ band (see Sect.~\ref{sub:rband}).

The \citet{ueda03} and \citet{laf05} HXLF predict very similar numbers
for the total counts, except for $z > 2$, where there are very few
constraining datapoints. In contrast, the \cite{aird10} HXLF
consistently predicts smaller values for the counts, especially at
fluxes greater than $10^{-13}$~erg~cm$^{-2}$~s$^{-1}$. This result is
because the \citet{aird10} HXLF predicts fewer high luminosity AGNs at all
redshifts. Moreover, the LADE model does not require an increase in
the space density of low-$z$, low-luminosity
AGNs. The predicted
number counts from the \citet{aird10} HXLF are different enough from
the other models that surveys will be able to test the viability of
this HXLF at different redshift ranges.

\subsubsection{Compton Thick AGNs}
\label{subsub:ctagns}
\begin{figure}
\plotone{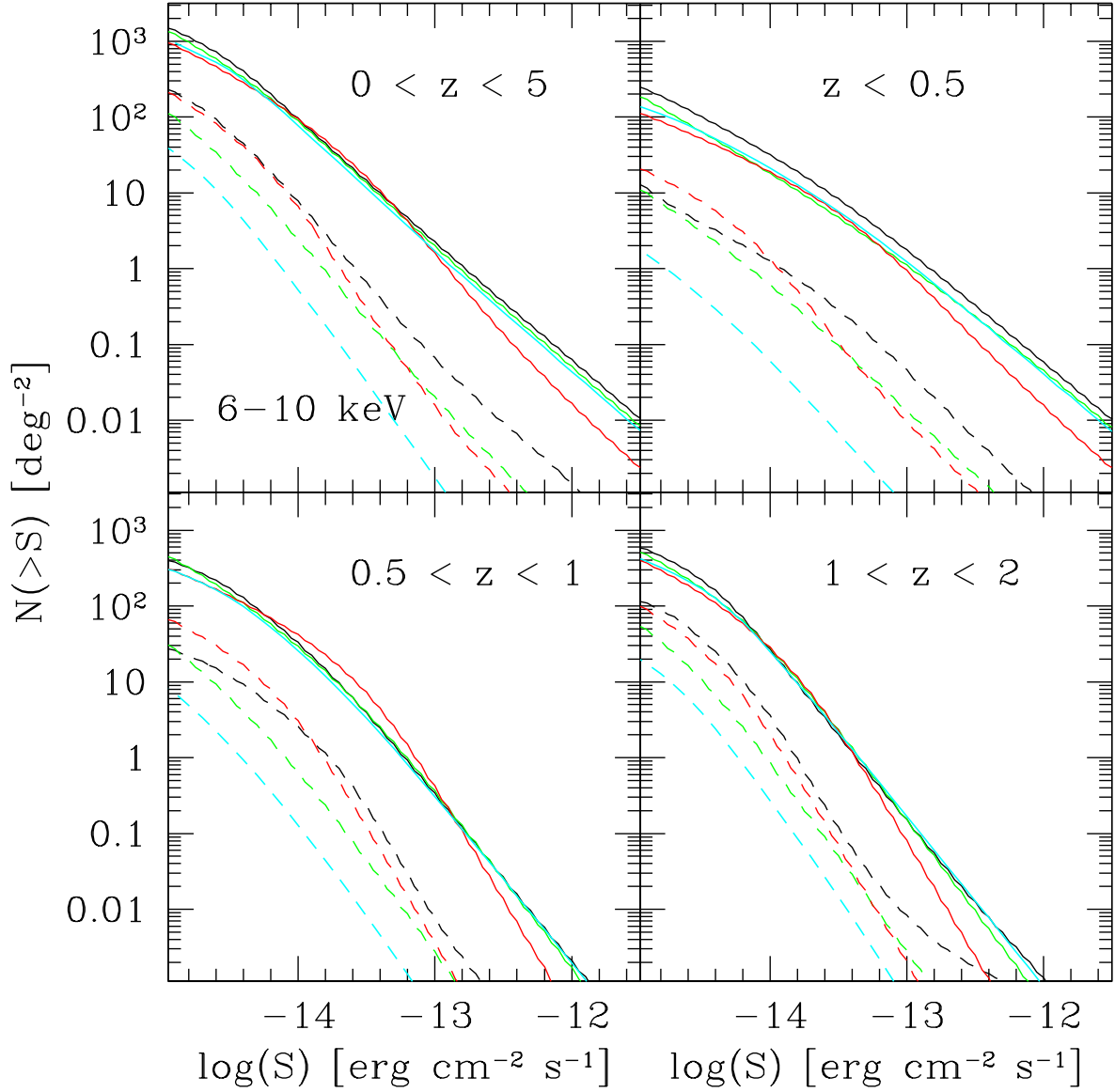}
\caption{The solid lines plot the integrated number counts of all AGNs
in the $6$--$10$~\kev\ energy band. The contribution to the counts by
Compton thick AGNs are shown as dashed lines. The top-left panel plots
the counts over the redshift range $z=0$ to $5$. The other panels show
the integrated counts over smaller ranges of redshift. The colors
differentiate between the various HXLFs and/or Compton thick
evolutions assumed by the model: \citet{laf05} HXLF (green), \citet{aird10} HXLF (red),
  the \citet{db10} composite model (black), and the conservative
  Compton thick model \citep{tuv09} (cyan). For clarity the model that uses the \citet{ueda03} HXLF has not been
  plotted.}
\label{fig:6to10counts}
\end{figure}

\begin{figure}
\plotone{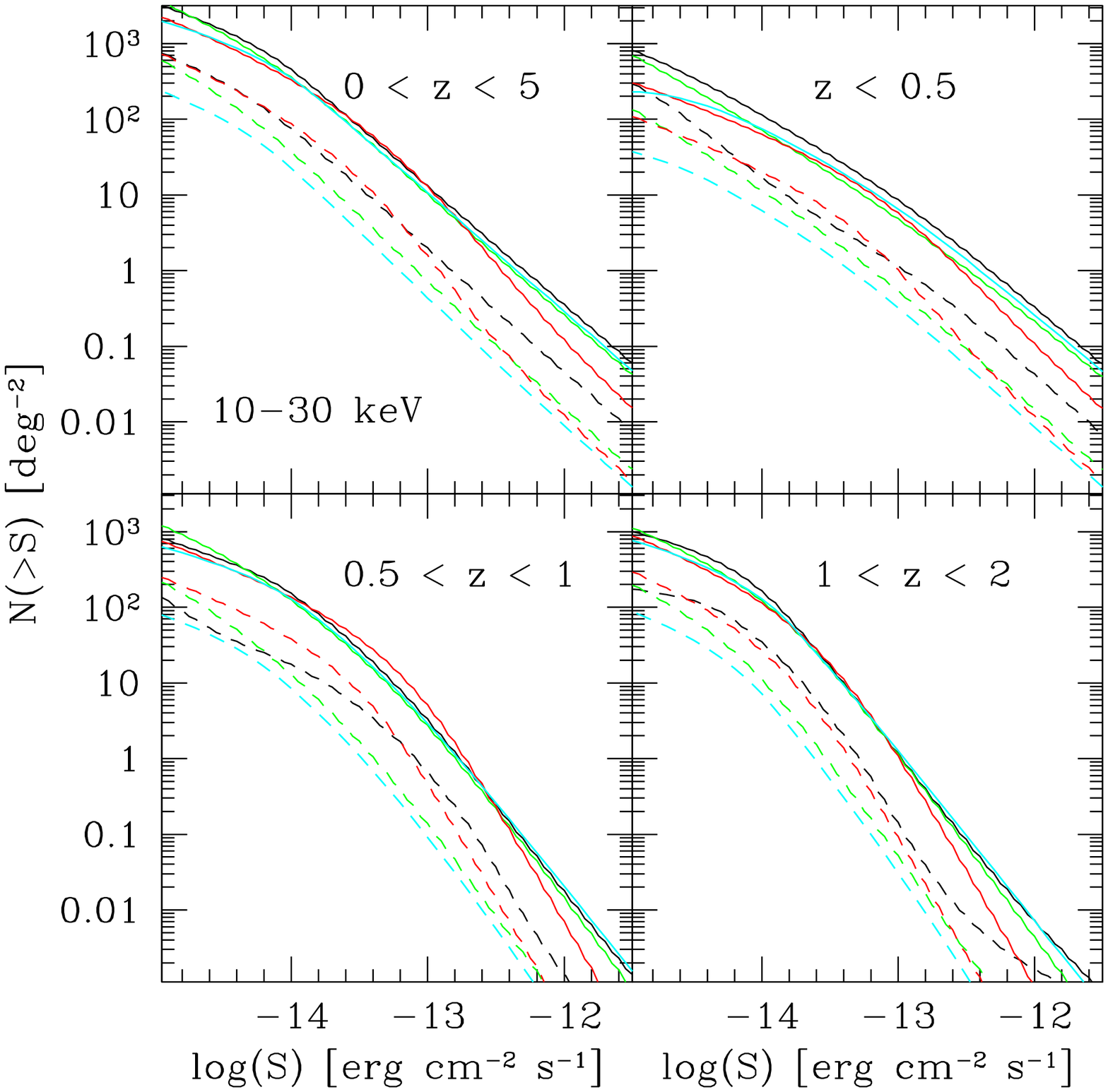}
\caption{As in Figure~\ref{fig:6to10counts}, but for the
  $10$--$30$~\kev\ energy band.}
\label{fig:10to30counts}
\end{figure}

\begin{figure}
\plotone{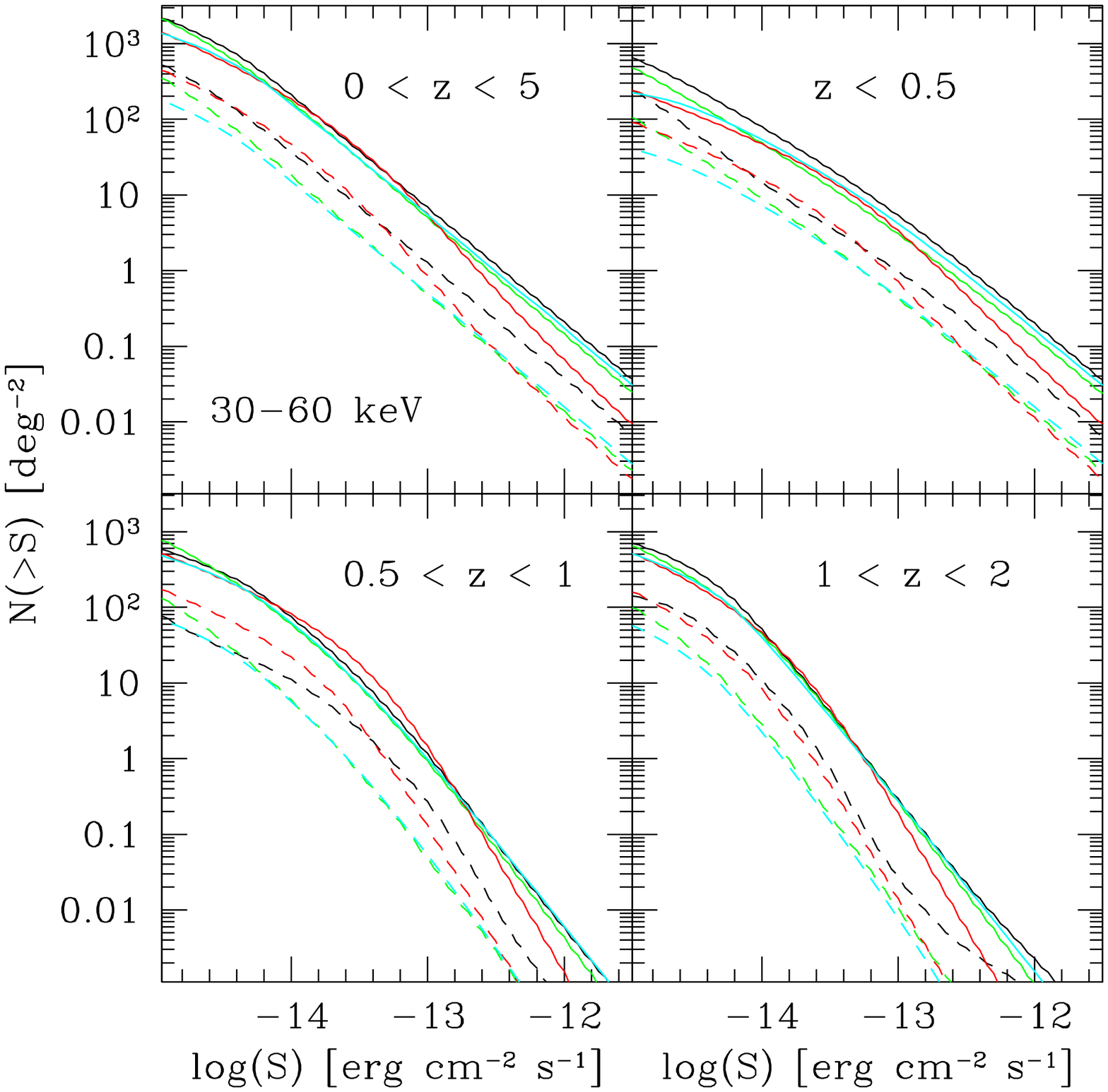}
\caption{As in Figure~\ref{fig:6to10counts}, but for the
  $30$--$60$~\kev\ energy band.}
\label{fig:30to60counts}
\end{figure}

Turning now to the Compton thick objects, we see that the number counts are
uniformly small, comprising only $5$--$10$\% of the total AGN counts. Therefore, discovering
and measuring the evolution of these objects will be a tough
observational challenge. However,
figures~\ref{fig:6to10counts}--\ref{fig:30to60counts} immediately show
the advantage of hard X-ray imaging to find Compton thick AGNs. At all
redshifts, the number counts of Compton thick source increase by factors of
$\sim10$--$100$ when moving from the $6$--$10$~\kev\ to the
$10$--$30$~\kev\ band. This result emphasizes that hard
X-ray imaging is an important method to identify significant
numbers of Compton thick candidates. Interestingly, there is little
difference in the expected Compton thick number counts between the $10$--$30$~\kev\
and $30$--$60$~\kev\ bands. In fact, the counts drop slightly in the
$30$--$60$~\kev\ band at $z \ga 1$ as the K-correction causes the
observed frame flux to drop. This is a result of the
the high-energy cutoff in the AGN power-law and electron recoil in
the reflection spectrum, both of which reduce the intensity of the
total spectrum at high energies. We conclude that, in order to
maximize the sensitivity to Compton thick AGNs, high energy observations
need only to extend out to $30$--$40$~\kev\ (depending on the redshift of
interest).

The combination of the different HXLFs and assumptions of the Compton
thick evolution result in wide range of predictions for the number
counts of these
AGNs. Figures~\ref{fig:6to10counts}--\ref{fig:30to60counts} indicate
that the best discrimination of the models will result from objects
at $z < 1$. In
particular, the model of \citet{db10} where Compton thick objects are
comprised of a combination of low Eddington ratio objects and very
high Eddington ratio AGNs (black lines), predicts a significantly different number
count distribution than the other four models where Compton thick objects are an
unchanging fraction of the AGN population, and evolve in the same way
as the less obscured sources. For example, at $z < 0.5$ the
\citet{db10} model predicts $\sim2\times$ more Compton thick AGNs with
a $10$--$30$~keV flux larger than $10^{-13}$~erg~cm$^{-2}$~s$^{-1}$ 
than the \citet{laf05} HXLF (Fig.~\ref{fig:10to30counts}). In the \citet{db10} model, these high
flux, low-$z$ Compton thick AGNs are dominated by low accretion rate objects similar
to the Circinus galaxy. Pushing to higher redshifts moves a survey
closer to the quasar era where high Eddington ratio accretion is more
common and the \citet{db10} model predicts additional Compton thick
sources. The factor between the evolving and fixed Eddington
ratio models at a depth of $10^{-13}$~erg~cm$^{-2}$~s$^{-1}$ is therefore
boosted to $\sim5$ in the $0.5 < z < 1$ bin. In addition, the evolving
Compton thick model has a shallower slope at bright fluxes than the
fixed Compton thick fraction models due to
low luminosity, weakly accreting Compton thick AGNs. Thus, the \citet{db10} model can be easily tested with a Compton thick AGN number counts
measurement that moves beyond $z \sim 0.1$ and to a $10$--$30$~keV
depth of $10^{-13}$~erg~cm$^{-2}$~s$^{-1}$. In this way hard X-ray
surveys will be able to measure the evolution of Compton thick AGNs
and determine their relation, if any, to stages of galaxy
evolution. Indeed, confirming that the Compton thick
AGN fraction is larger at higher redshift will have implications for
understanding the physics of AGN triggering and the AGN-galaxy
connection. Finally, as seen by comparing the conservative
\citet{tuv09} model (cyan lines) to the other predictions, the number of Compton thick AGNs detected
by hard X-ray surveys will help test the assumptions made in that
model, in particular the shape of the average spectral model and the
value of the local Compton thick density. 

\subsection{Identification and Optical Detectability of the Hard X-ray Sources}
\label{sub:rband}
\begin{figure}
\plottwo{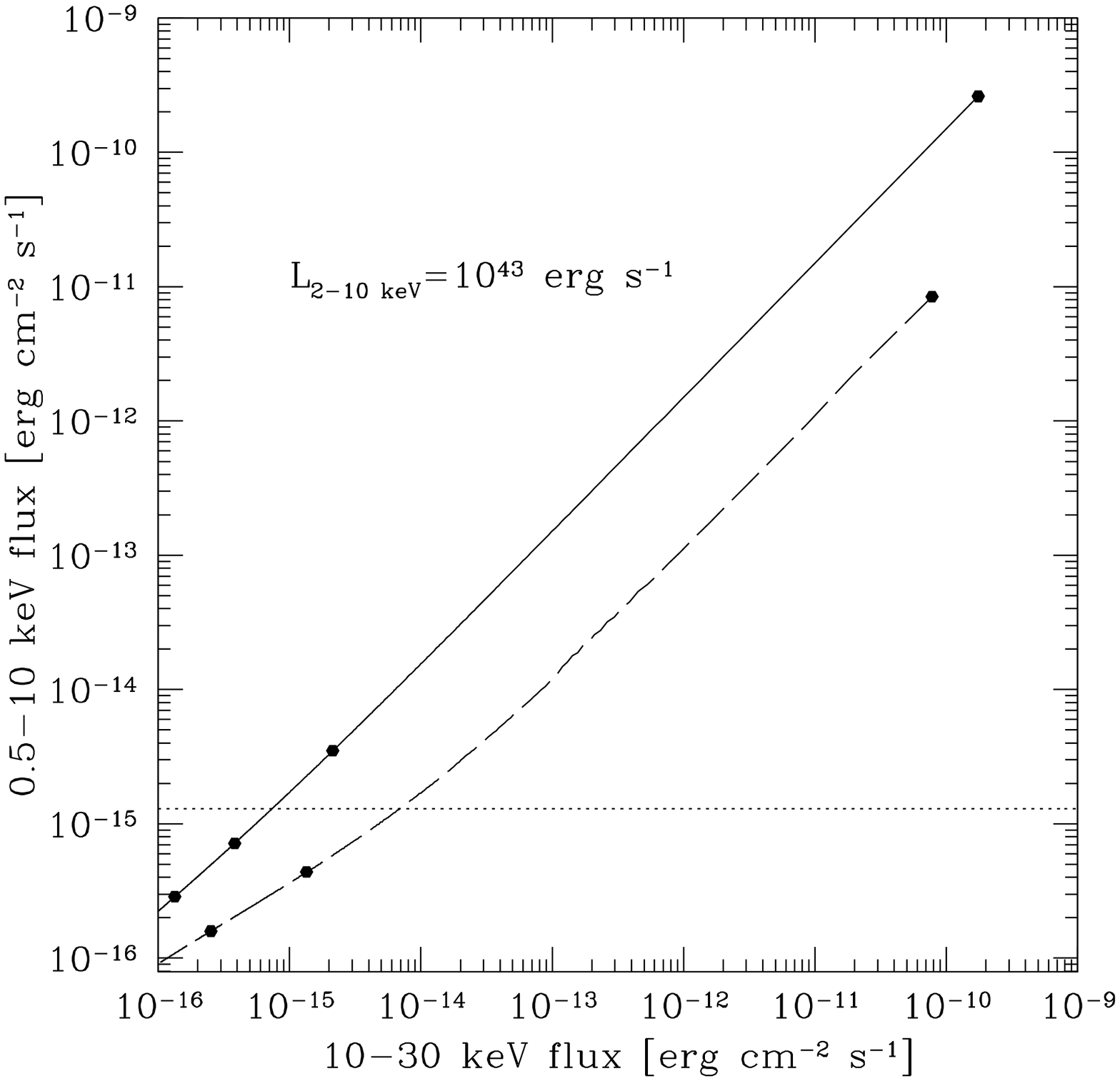}{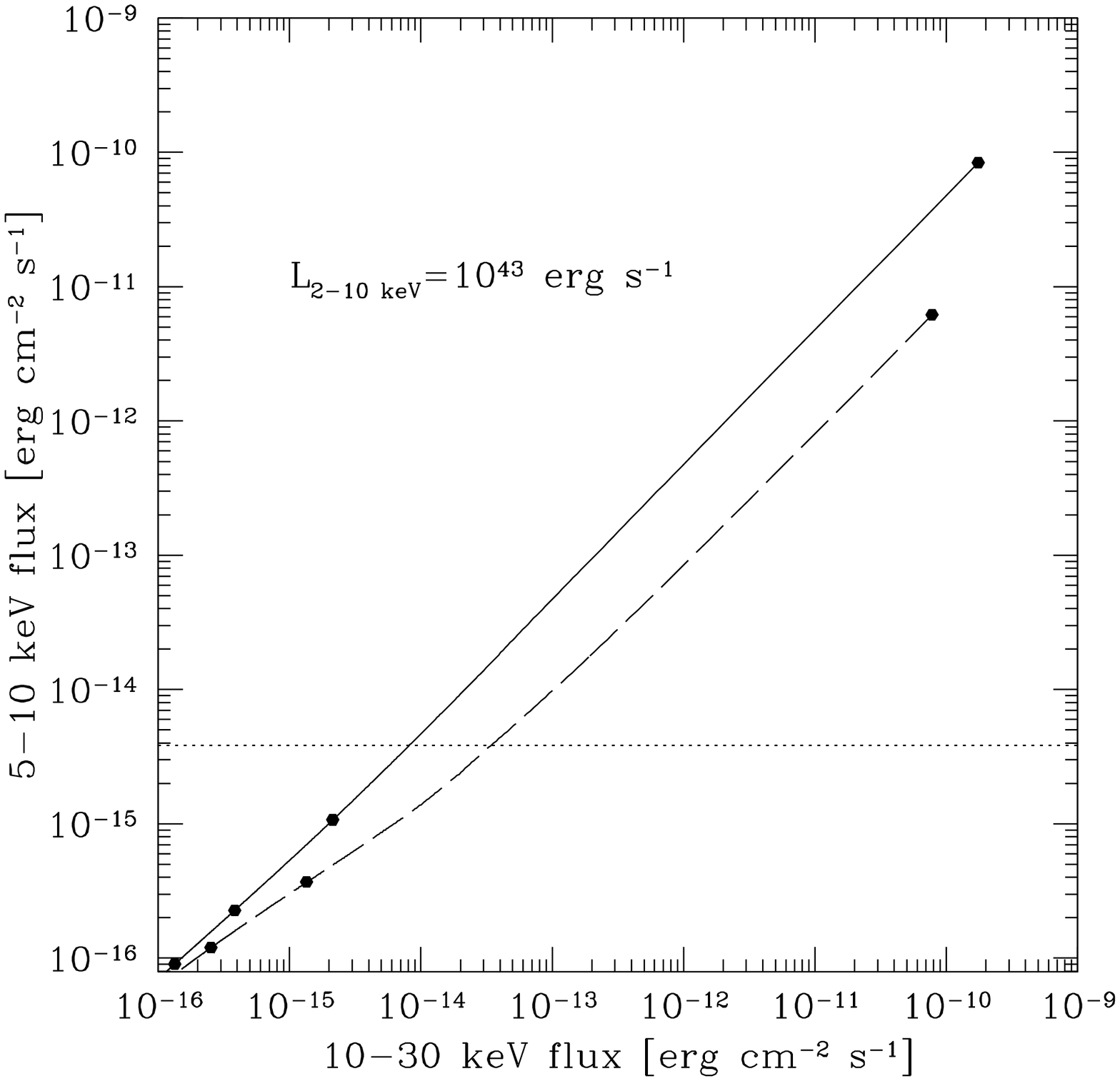}
\caption{Expected fluxes in the $0.5$--$10$~keV band (left) and the
  $5$-$10$~\kev\ band (right) as a function of the $10$--$30$~keV fluxes for AGNs with intrinsic, 
unabsorbed $2$--$10$~keV luminosities of $L_X=10^{43}$~erg~s$^{-1}$. The solid line
plots the fluxes for AGNs with $N_H=10^{21-21.5}$~cm$^{-2}$ and the
dashed line indicate fluxes for AGNs with
$N_H=10^{24-24.5}$~cm$^{-2}$. Redshift increases from upper 
right to lower left; the filled circles show the redshifts for each
track at $z=0$, $1$, $2$ and $3$ from right to left respectively. The
  horizontal lines indicate the flux limit of the \chandra\
  survey (eight 200~ks exposures) of the AEGIS field to which 50\% of the total survey area is
  complete \citep{lai09}.}
\label{fig:flux_chandra_nustar}
\end{figure}

As seen above, hard X-ray imaging is extremely efficient in detecting
AGNs, but additional analysis will be required in order to begin to
use these detections to investigate problems in galaxy and black hole
evolution. This is particularly true for the Compton thick AGNs, as
they cannot be identified from hard X-ray imaging alone --- additional
data is necessary to determine the level of obscuration for each hard
X-ray source. Moreover, understanding the nature of the detected hard X-ray sources will
require that the host galaxies be sufficiently bright so that redshifts
and stellar population measurements can be made at optical and
infrared wavelengths. Both of these problems can be addressed if the
hard X-ray surveys are performed on fields with significant
multiwavelength coverage. For example, we have seen that the majority of hard X-ray sources
should have \chandra\ or \xmm\ counterparts provided that at least
moderately deep (i.e., a flux limit of $\sim10^{-16}$--$10^{-15}$~erg~cm$^{-2}$~s$^{-1}$) soft X-ray observations exist in the observed
field(s). Figure~\ref{fig:flux_chandra_nustar} shows the relationship
between the soft and hard X-ray fluxes for the spectral model
described in Sect.~\ref{subsub:hxlfmods}. The left panel plots $0.5$--$10$~\kev\ fluxes for
AGNs with (intrinsic, unabsorbed) $L_X=10^{43}$~erg~s$^{-1}$ against
the $10$--$30$~\kev\ flux for different obscuring column
densities. As an example of the typical \chandra\ sensitivity, the
horizontal line shows the 50\%-completeness flux limit from the
\chandra\ observations of the All-wavelength Extended Groth Strip
International Survey (AEGIS, eight 200~ks exposures; \citealt{lai09}). The solid line shows the fluxes
for unabsorbed AGNs, and, for all realistic flux
limits reached by the next generation of hard X-ray surveys
($\sim10^{-14}$--$10^{-13}$~erg~cm$^{-2}$~s$^{-1}$ in the $10$--$30$
keV band), these sources will have $0.5$--$10$~keV fluxes greater than
the flux limit for moderate/deep \chandra\ observations in many survey
regions. Thus, $\sim 90$\% of the hard X-ray detections will have soft
X-ray counterparts that can be combined with the hard X-ray data to
estimate $N_{\mathrm{H}}$. 

The dashed line plots the fluxes for Compton
thick AGNs, and the effects of absorption are clearly seen
in the predicted $0.5$--$10$~\kev\ fluxes. Although Compton thick AGNs
will be an order of magnitude brighter in the $10$--$30$~\kev\ band,
ones detected in the hard X-rays still seem to be at or above the detectability limit of the
deepest \chandra\ surveys (if they lie at low $z$). However, this plot
neglects the steep decline in the ACIS-I effective area at energies
greater than $5$~\kev. This effect is illustrated in the right panel
of Fig.~\ref{fig:flux_chandra_nustar} which plots the $5$--$10$~\kev\
flux versus the $10$--$30$~\kev\ flux. This figure clearly shows that some Compton
thick AGNs will be detectable in hard X-ray surveys, but beneath the
sensitivity limit of deep \chandra\ surveys. These results again indicate the power of
hard X-ray imaging, with no decline in effective area above 5~\kev, to
discover the most heavily obscured AGNs between $z=0$ and $1$. Thus,
in addition to those that might be found by combining \chandra\ and
hard X-ray detections, Compton thick candidates can also be identified by searching for objects
that have no $5$--$10$~\kev\ \chandra\ counterparts. Analysis using
infrared data \citep[e.g.,][]{alex08} can then be pursued on these
candidates to confirm the identification. Finally, there may be an
unknown number of heavily obscured objects that have extremely weak
scattered components in the soft band \citep[e.g.,][]{ueda07}. Such
objects would also be detected out to moderate redshift by hard X-ray
imaging.

Redshifts are typically obtained from spectroscopic or photometric
measurements of the host galaxy. Therefore, it is important to
consider the expected $R$-band magnitudes for
different $10$--$30$~\kev\ fluxes to obtain a sense of the difficulty
of identifying the host galaxy of the hard X-ray detected AGNs. This is
done by using the empirical relations between X-ray luminosity and
absolute $R$-band magnitude reported by \citet{laf05}, measured
independently for obscured and unobscured AGN:
 
\begin{equation}
\log L_R=(0.959 \pm 0.025) \log L_X + (2.2 \pm 1.1),
\end{equation}

\noindent
for unobscured AGN, and

\begin{equation}
\log L_R=(0.462 \pm 0.026) \log L_X + (23.7 \pm 1.1)
\end{equation}

\noindent
for obscured ones, where $L_X$ is the intrinsic luminosity in the
rest-frame $2$--$10$~keV band. These equations are obviously an
approximation, as they do not take into account some of the more
detailed properties of the host galaxies, like evolving stellar
populations or varying amounts of extinction\footnote{However, the
\citet{laf05} sample has a similar $z$ distribution as will be found
in the next generation of hard X-ray surveys, so the observed evolution of $R$-band
luminosity should be similar.}. Such considerations are
particularly important for obscured and Compton thick AGN, in which
the optical light is dominated by the host galaxy. The results
are indistinguishable from the calculations, based on a dusty torus
model, by \citet{treister04}, which assumed the same host galaxy for
all AGN.

For the deepest hard X-ray surveys with an limiting $10$--$30$~\kev\ sensitivity of
$10^{-14}$~erg~cm$^{-2}$~s$^{-1}$, most AGNs will have bright
optical magnitudes, $R<23$ (AB), and, thus for surveys in well studied
fields, the vast majority of the sources will
have detections at other wavelengths and measured redshifts. For a
very shallow X-ray survey ($\sim 10^{-13}$~erg~cm$^{-2}$~s$^{-1}$
depth), most AGNs will be brighter than $R \sim 20$ mag (AB), and hence
within reach of the Sloan Digital Sky Survey.

\section{Survey Strategies for Hard X-ray Missions}
\label{sect:surveys}
\subsection{Area vs. Depth}
\label{sub:areavsdepth}
\begin{figure}
\epsscale{0.74}
\plotone{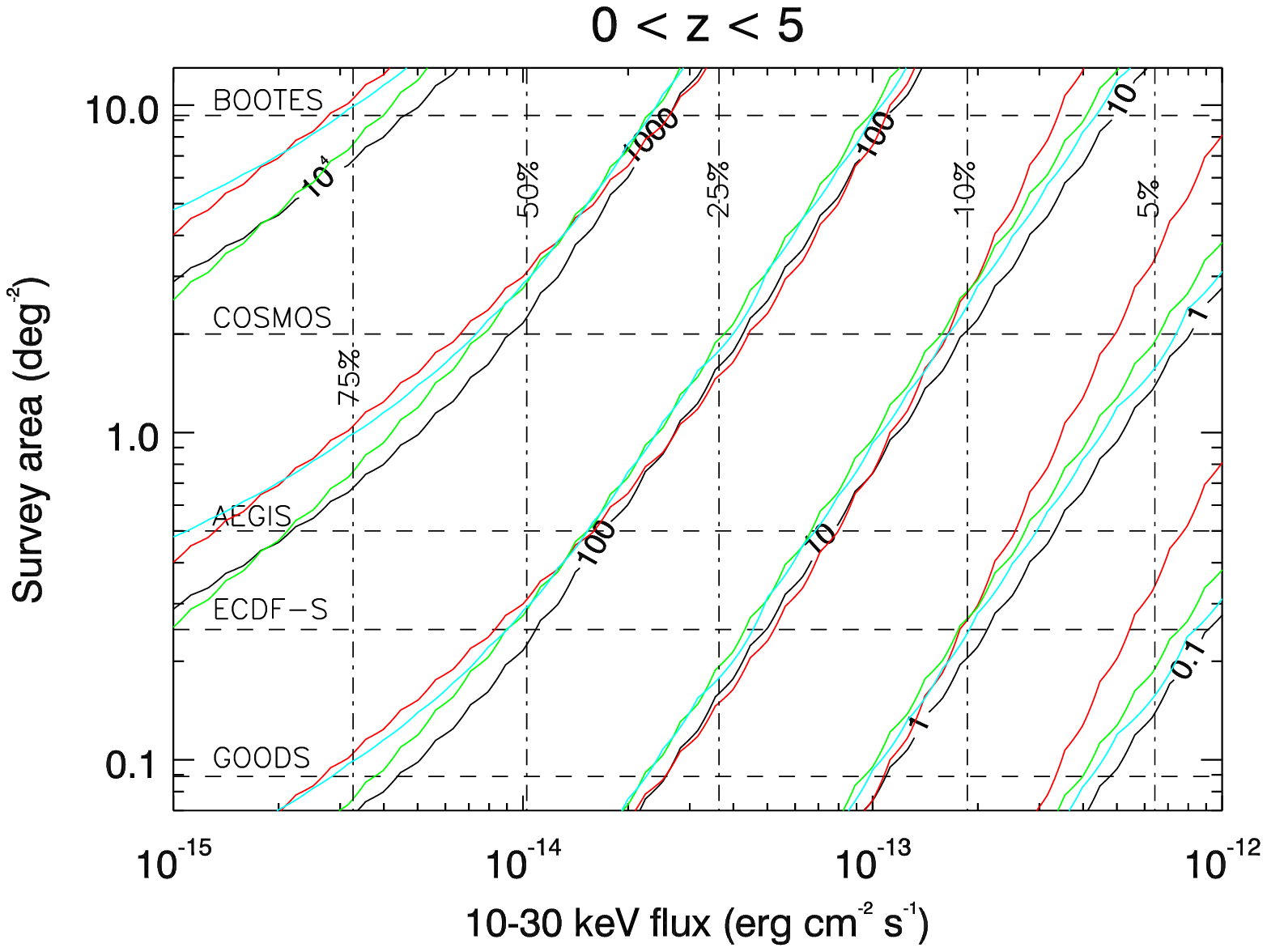}
\plotone{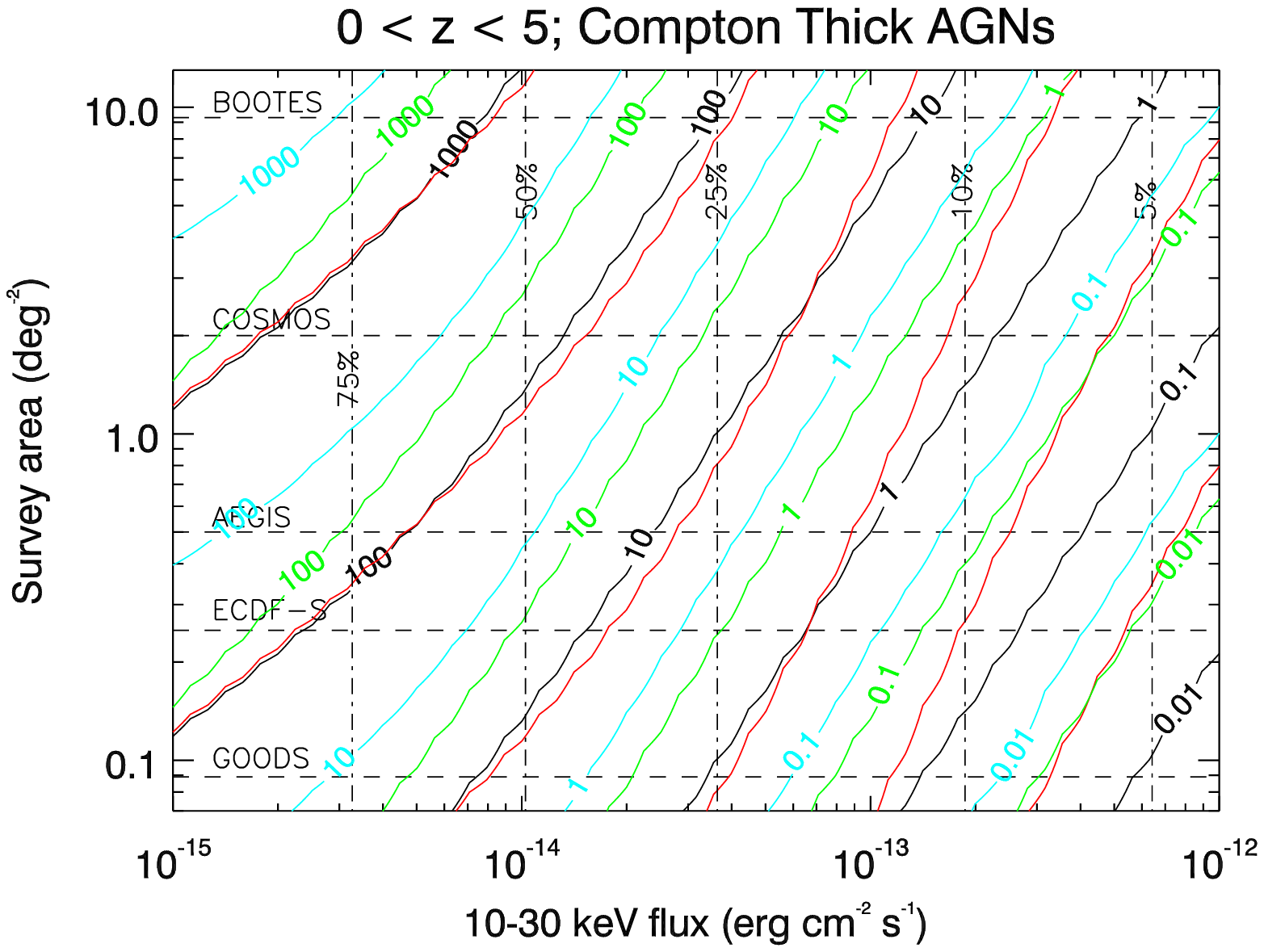}
\caption{(Top) Contours of number of AGNs as a function of survey area and
  $10$--$30$~\kev\ sensitivity. Results for the models are plotted
  using the same color scheme as previous figures with the
  \citet{ueda03} HXLF omitted for clarity. These contours include all sources in
  the range $0 < z \leq 5$. The contour levels increase by
  factors of 10 from 0.1 sources (lower-right corner of the figure) to
  10,000 sources (upper-left corner of the figure). Only the black contours are labeled. The horizontal lines show the
areas covered by some well known multiwavelength surveys. The vertical
lines indicate the flux level required to reach a certain percentage
of the X-ray background (as judged by the \citealt{ueda03} model) in
the $10$--$30$~\kev\ band. (Bottom) Contours of Compton thick AGNs. The
  contour levels increase by factors of 10 from 0.01 sources (lower-right corner of the figure) to
  1000 sources (upper-left corner of the figure).}
\label{fig:contour10to30}
\end{figure}

\begin{figure}
\plotone{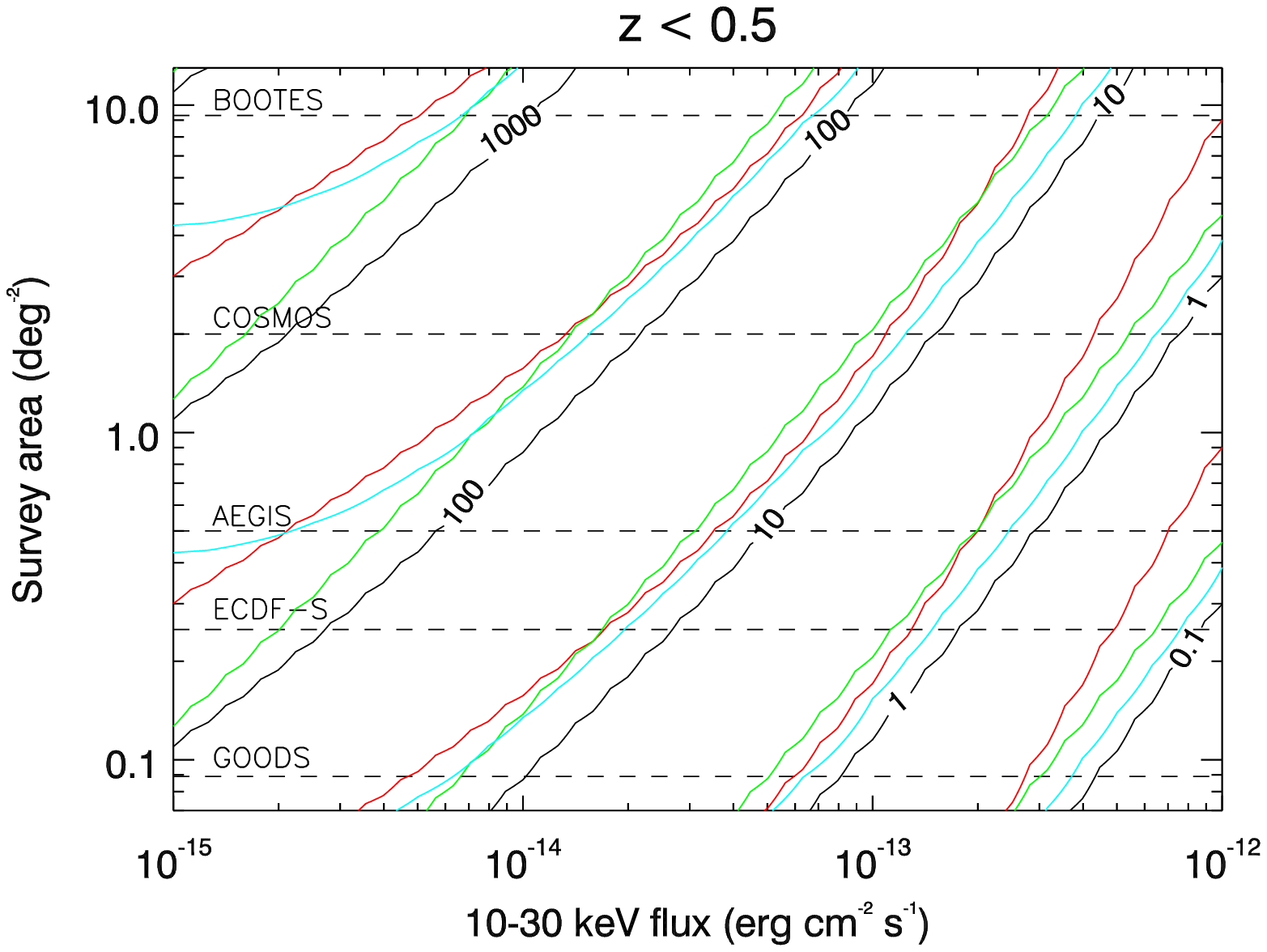}
\plotone{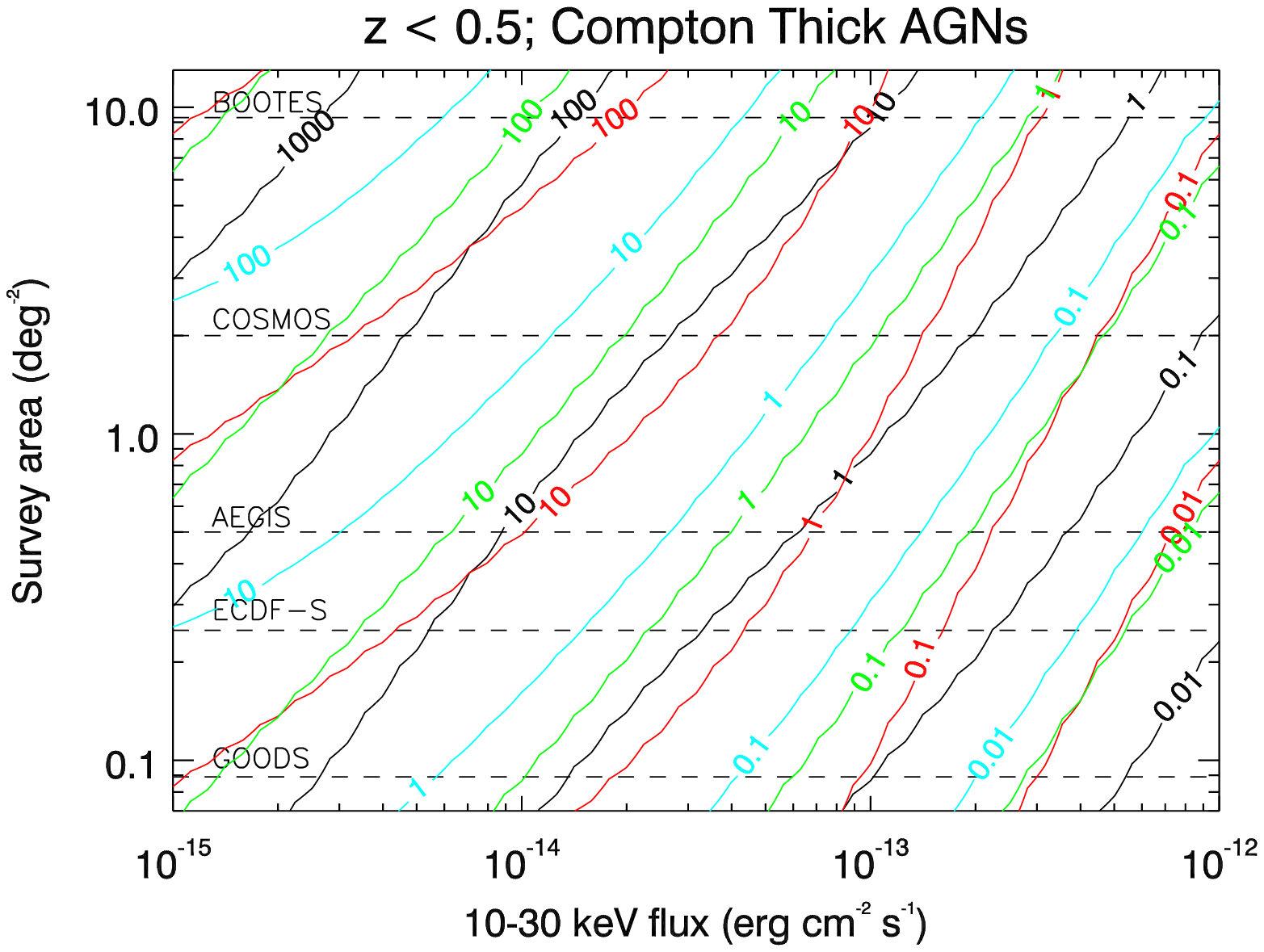}
\caption{As in Fig.~\ref{fig:contour10to30}, however these contours
  include only those sources in
  the range $0 < z < 0.5$.}
\label{fig:contour10to30kevz05}
\end{figure}

Deep extragalactic surveys require significant investments of
observing time and so need to be carefully planned in order to
maximize the scientific return. One important consideration in survey
planning is the trade-off between depth and area, as the luminosities
and redshifts of sources detected in a very deep and narrow field is
often very different from those uncovered in a shallower but wider
field. The predicted number counts shown in
Figs.~\ref{fig:6to10counts}--\ref{fig:30to60counts} contain all the
information necessary for survey planning, but require additional
manipulation to see the effects of surveys of differing
areas. Therefore,
Figs.~\ref{fig:contour10to30} and~\ref{fig:contour10to30kevz05}
present examples of a new way of plotting the predicted AGN and Compton thick AGN counts
that is specifically designed for planning hard X-ray surveys. Each
figure plots contours of numbers of AGNs (upper panel for all AGNs;
lower panel for Compton thick AGNs) as a function of survey area and
$10$--$30$~\kev\ flux. The different colors
indicate the predictions for four of the five different models considered in
this paper (see Sect.~\ref{sub:models}). The dashed horizontal lines in
each plot show the area of five regions with good
multiwavelength coverage: Bo\"otes Deep Wide-Field Survey
(9.3~deg$^2$; \citealt{murr05}), COSMOS (2~deg$^2$; \citealt{has07,elv09}),
AEGIS (0.5~deg$^2$; \citealt{dav07,lai09}), the Extended Chandra Deep Field South
(ECDF-S, 0.25~deg$^2$; \citealt{leh05}), and the Great Observatories
Origins Deep Survey (GOODS, 0.089~deg$^2$;
\citealt{alex03,luo08}). Finally, both panels of
figure~\ref{fig:contour10to30} also have vertical lines indicating the flux needed
(estimated using the \citealt{ueda03} model) to reach a specific
percentage of the integrated $10$--$30$~keV X-ray background spectrum.

These figures contain a large amount of information and should be
useful guides for planning future hard X-ray surveys\footnote{Plots for
different energy bands and redshift ranges are available by contacting
the authors.}. Close
examination of the contours shows that, with the exception of the
\citet{aird10} HXLF, all the models predict very similar numbers for
the total number of AGNs. There is a slight
spreading out of the contours at small fluxes, illustrating the
uncertainty in the slope and evolution of the faint end of the
HXLF. In contrast, there is a significant difference in the number of
Compton thick AGNs predicted at a given flux and survey area,
indicating that deep surveys will be able to easily distinguish between
the various models.

Identifying highly embedded AGNs over a range of redshifts will
be one of the principle goals of any hard X-ray survey; however, these
AGNs will be faint and will require a significant investment of
observing time to obtain a sizable number of sources. Fig.~\ref{fig:contour10to30kevz05} shows
that an area the size of the COSMOS field must be surveyed to a depth
of $2$--$3\times 10^{-14}$~erg~cm$^{-2}$~s$^{-1}$ in the
$10$--$30$~\kev\ band to uncover at most 10 Compton thick AGNs at $z <
0.5$. Similar numbers of Compton thick sources would also be found for
that depth and sensitivity at
$0.5 < z < 1$  and $1 < z < 2$. In contrast, a narrow field
(such as the GOODS field) would have to be observed to a flux limit
$10\times$ fainter to obtain similar numbers of $z > 0.5$
Compton thick AGNs. This survey, however, would not detect as many $z
< 0.5$ objects.  

\subsection{Application to \nustar}
\label{sub:nustar}
As mentioned in Sect.~\ref{sect:intro}, only a portion of \nustar's
baseline two year mission will be dedicated to extragalactic surveys. The limited
amount of time that is available requires that careful pre-launch
planning be performed to optimize the survey strategy. Here we make use
of the hard X-ray number counts presented above to predict the results
of a \nustar\ survey of five different regions with significant
ancillary multiwavelength data, including deep low-energy X-ray observations: Bo\"{o}tes, COSMOS,
AEGIS, ECDF-S and GOODS. Each survey has a total exposure time of
6.2~Ms, which, assuming 50\% efficiency, would take 6 months to complete.  

The \nustar\ sensitivities were derived using the ``nustar\_sens''
program provided by the \nustar\ instrument team. The program
incorporates the \nustar\ effective area, corrected for aperture stop,
scattering from surface roughness, figure error, detector efficiency,
and all known attenuation along the optical path in the form of
thermal covers and beryllium windows. Combined with the GEANT4 \citep{ago03}
simulated internal background and the diffuse X-ray background (for a
low Earth orbit inclined at 5$\degr$), the
program calculates a sensitivity map for the entire detector given a
specific signal to noise, point spread function (PSF) and input
spectrum. A $\Gamma=1.7$ spectrum, appropriate for a moderately
obscured AGN uncorrected for absorption and reflection \citep[e.g.,][]{malizia03,winter09}, is assumed for all calculations, and
we require a $4\sigma$ detection. 
Since the code was run for long exposures the signal to noise is in
the Gaussian regime. The PSF used for the calculations follows a
Gaussian+King profile and has an half-power diameter (HPD) of
$41$\arcsec. Because extraction from a region larger than the HPD will
add more background than source counts, the HPD is used as the
extraction region. 

By definition, surveys require mapping areas greater than the
detector field-of-view ($13^{\prime} \times 13^{\prime}$), so survey sensitivities were calculated for
two tiling strategies. The first, called `corner shift', is when the
survey region is covered by non-overlapping fields of view with
additional observations at the corners between fields of view. This strategy has the
advantage that individual exposures can be fairly long, but there is a
significant variation in sensitivity over the survey region as
\nustar, like many other high-energy instruments, has an effective
area that is a strong function of off-axis angle. The second strategy
is called `half shift', and was calculated by shifting the detector sensitivity map over by half a field of
view in both directions. Each individual pointing is shallower than
the corner shift, but
the half shift provides a more uniform sensitivity map over the survey
region. One by-product of the half shift survey is that, due to the fact that the
pointings are calculated such that they entirely cover the given
survey size with the best sensitivity, the actual survey size ends up
being larger than the input, with the extraneous area having
lower sensitivity.

\begin{figure}
\plotone{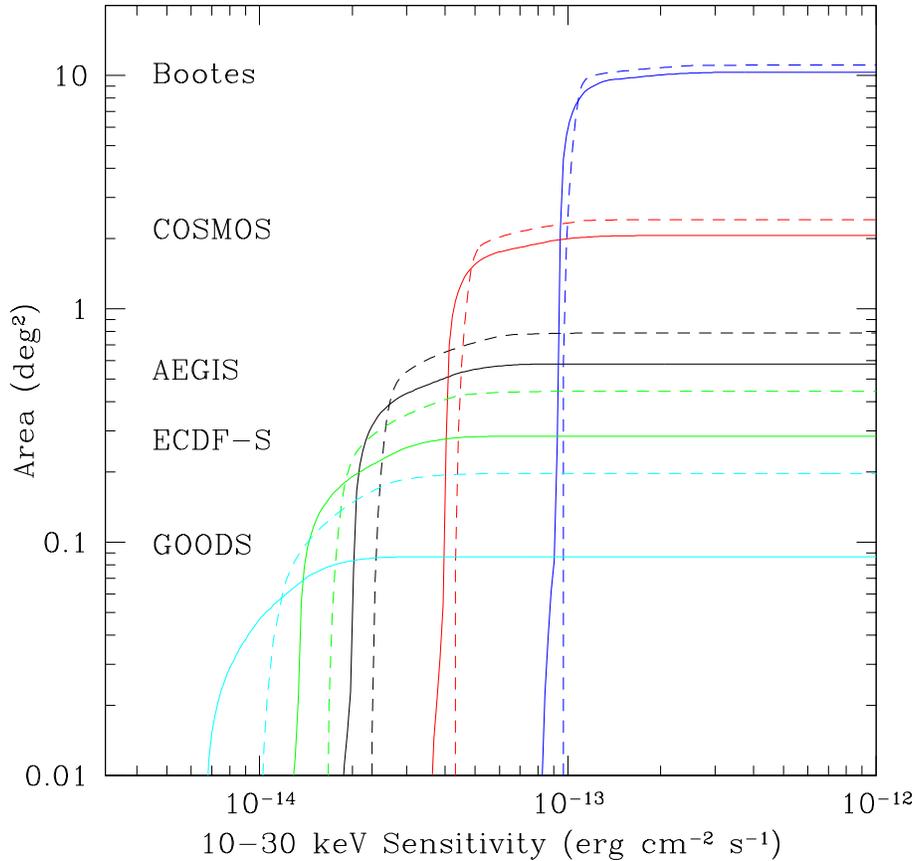}
\caption{Area versus $10$--$30$~\kev\ sensitivity for 4$\sigma$
  detections of AGNs in different \nustar\ surveys with a total
  exposure time of 6.2~Ms (about 6 months, assuming 50\%
  efficiency). The solid lines are for a corner shift survey, and the
  dashed lines plot the results for a half shift survey. Results are
  shown for surveying the following regions: Bo\"{o}tes (blue), COSMOS
  (red), AEGIS (black), E-CDFS (green) and GOODS (cyan).}
\label{fig:surveysens}
\end{figure}

The area versus sensitivity curves for the five surveys are shown in
Figure~\ref{fig:surveysens} with the corner shift surveys plotted as
solid lines and the half shift surveys shown as dashed lines. The
details of each survey --- including the number of pointings, average
sensitivity, and resolved fraction of the X-ray background --- are
listed in Table~\ref{table:cornershift}. As expected, the corner shift
surveys probe slightly deeper than the half shift ones, but the latter
provide more uniform coverage over the survey area, especially in the
narrow-fields like GOODS or ECDF-S. The numbers of AGNs predicted from
both tiling strategies are nearly equal, but the wider area provided by
the half shift survey increases the AGN yield by $\sim 10$\%. Thus, we conclude that half shift
tiling should be used for all \nustar\ surveys.

Table~\ref{table:halfnum} shows the number of AGN detections predicted
by each of the five XRB models for each of the five half shift
\nustar\ surveys. The significant decrease in effective area with
energy is manifested in the numbers of detectable AGN: the majority of
sources will be detected in the $6$--$10$~\kev\ band with the
$30$--$60$~\kev\ band yielding very few objects and basically no
Compton thick objects. However, $\sim 100$ AGNs will be detected in
the $10$--$30$~\kev\ including (depending on the model) $\sim
10$--$20$ Compton thick sources. Moreover, these objects will be
detectable up to and beyond $z \sim 2$. These surveys will therefore
increase the number of AGNs detected in hard X-rays by over an order
of magnitude, and will be able to estimate the evolution of Compton
thick AGNs above $z \sim 0$. In addition, good spectral
measurements of absorbing column density, reflection strength and
high energy cutoff will be determined for a large number of AGNs over
a range of luminosity and $z$. Such measurements were only
previously available for a small number of nearby objects
\citep{beck09,mol09}. Interestingly, because of the higher effective area and
the negative K-correction, most Compton thick AGNs will also be
detected in the $6$--$10$~\kev\ band unless the exposures are very
shallow (e.g., see the numbers for Bo\"{o}tes field in Table~\ref{table:halfnum}). As the \nustar\ effective
area peaks at $\sim 10$~\kev, some fraction of the low-$z$ heavily obscured AGNs will not be
detected by the deep \chandra\ surveys (see Fig.~\ref{fig:flux_chandra_nustar}).


The choice of \nustar\ surveys must be driven by the science goals:
discovering the sources that contribute to the peak of the XRB, and
determining the evolution of highly obscured sources. The predicted number of Compton thick detections is
roughly constant for all the surveys (except for the narrowest
GOODS-like field), with the redshift
distribution shifting to higher redshift as the field
narrows. This fact allows some flexibility in designing a survey; for
example, a 6.2~Ms survey with an area $\sim 2$~deg$^2$ will detect $15$--$20$ Compton thick AGNs with redshifts $z \la 2$. Such a
survey will also resolve $\sim 30$\% of the XRB in $10$--$30$~\kev\
band. The variation in predicted numbers of Compton thick objects
implies that a simple counting experiment that can push to $z \sim 1$ may provide important
constraints on the models. A very deep field such as the ECDF-S will
also be of value, as it will probe a different part of the $z$-luminosity plane, and therefore would increase the
statistics of the high-$z$, highly obscured population, and improve 
the resolved fraction of the XRB at high energies by detecting a
larger number of lower luminosity AGNs at high $z$.

Finally, recall that the numbers discussed here assume that 6.2~Ms are
set aside
for extragalactic surveys in the baseline \nustar\ two-year
mission. However, the spacecraft has no consumables and is expected to
have an orbital lifetime greater than 5 years
\citep{harr10}. An extended \nustar\ mission will be able to spend
several months per year building up and extending the surveys
performed during the baseline mission. Thus, \nustar\ may end its
mission with extragalactic surveys that have exposure times 2 to 3
times larger than the 6.2~Ms assumed here, with the numbers of
detected AGNs correspondingly larger.

\section{Conclusions}
\label{sect:concl}
The next few years will see the launch of the first generation of hard
X-ray focusing telescopes. By focusing X-rays with energies $\ga
10$~\kev\ these missions will increase by orders of magnitude the
sensitivity of observations in this energy range. This capability will
be vitally important to study the growth of accreting black holes, as
much of this accretion is expected to be obscured by significant
columns of gas and dust. This paper presents predictions for AGN
number counts in three hard X-ray bands ($6$--$10$~\kev,
$10$--$30$~\kev, and $30$--$60$~\kev) to understand and quantify the
potential of extragalactic surveys performed by these future
missions\footnote{The results can easily be extended to other energy
and flux ranges by contacting the authors.}. The calculations pay
particular attention to the expected number of Compton thick AGNs, as
these heavily embedded objects may be physically connected to rapid
galaxy growth and are missed by previous observations. We
present predictions for five different models of the XRB and Compton
thick evolution. Specifically, results are presented for three
different measurements of the AGN HXLF which differ significantly in
the predicted evolution of high-$z$ low-luminosity AGNs. A fourth model
assumes the Compton thick fraction is a function of the AGN Eddington
ratio. The final model presents the most conservative view of the Compton thick
population. These five models predict numbers of Compton thick AGNs
that differ up to a factor of ten, therefore any measurement of the
Compton thick number counts beyond $z=0$ will highly constrain the
fraction and evolution of Compton thick AGNs.

Most hard X-ray detected AGNs will be \chandra\ or \xmm\ sources and
have host galaxies with $R < 23$; therefore, sources which are well separated
from the background should be easily identified and have counterparts in
multiwavelength ancillary data. Combining the hard and soft X-ray data
will also allow the $N_{\mathrm{H}}$ estimates that are are necessary
to identify Compton thick AGNs. Other Compton thick candidates will be
identified by following up sources that are undetected in the
\chandra\ or \xmm\ data. In order to make use of these data, hard X-ray surveys should be
performed in well-observed regions with deep and uniform
multiwavelength coverage.

Specific predictions for five extragalactic surveys were performed for the
\nustar\ mission, scheduled for launch in early 2012. All the surveys
assumed a total available exposure time of 6.2~Ms, and a half shift
tiling strategy is recommended. These deep surveys will yield hundreds
of AGNs at $z \la 2$, including (depending on the area covered)
$10$--$20$ Compton thick AGNs over that redshift range. Thus, the
\nustar\ deep surveys will allow a detailed exploration of the
nature and evolution of nuclear obscuration in galaxies. The
simulations were appropriate for the two-year baseline mission of
\nustar, but the surveys will ultimately probe much deeper during the
extended \nustar\ mission.

The predictions presented here show that, assuming equal
sensitivities, hard X-ray surveys can more
efficiently detect AGNs and follow their evolution than lower energy
X-ray observations. It 
is therefore hoped that future technology development will increase the
sensitivity of hard X-ray instrumentation. A future highly efficient
hard X-ray imaging mission, in combination with a ground based 30m
optical/IR telescope, \textit{JWST} and ALMA, would be extraordinarily
powerful in understanding galaxy and black hole growth throughout
cosmic time.
  
\acknowledgments
The author list is alphabetical; all authors contributed equally to
this work. This work was supported in part by NSF award AST 1008067 to
DRB. JR was supported by a Carnegie Fellowship at the Carnegie
Observatories in Pasadena. Support for the work of ET was provided by the National
Aeronautics and Space Administration through \chandra\ Post-doctoral
Fellowship Award Number PF8-90055 issued by the \chandra\ X-ray
Observatory Center, which is operated by the Smithsonian Astrophysical
Observatory for and on behalf of the National Aeronautics Space
Administration under contract NAS8-03060. The authors thank D.\
Alexander, F.\ Harrison and D.\ Stern for comments.

\appendix
\section{Predicted $5$--$10$~keV Counts}
\label{sect:app1}
\begin{figure}
\figurenum{A1}
\plotone{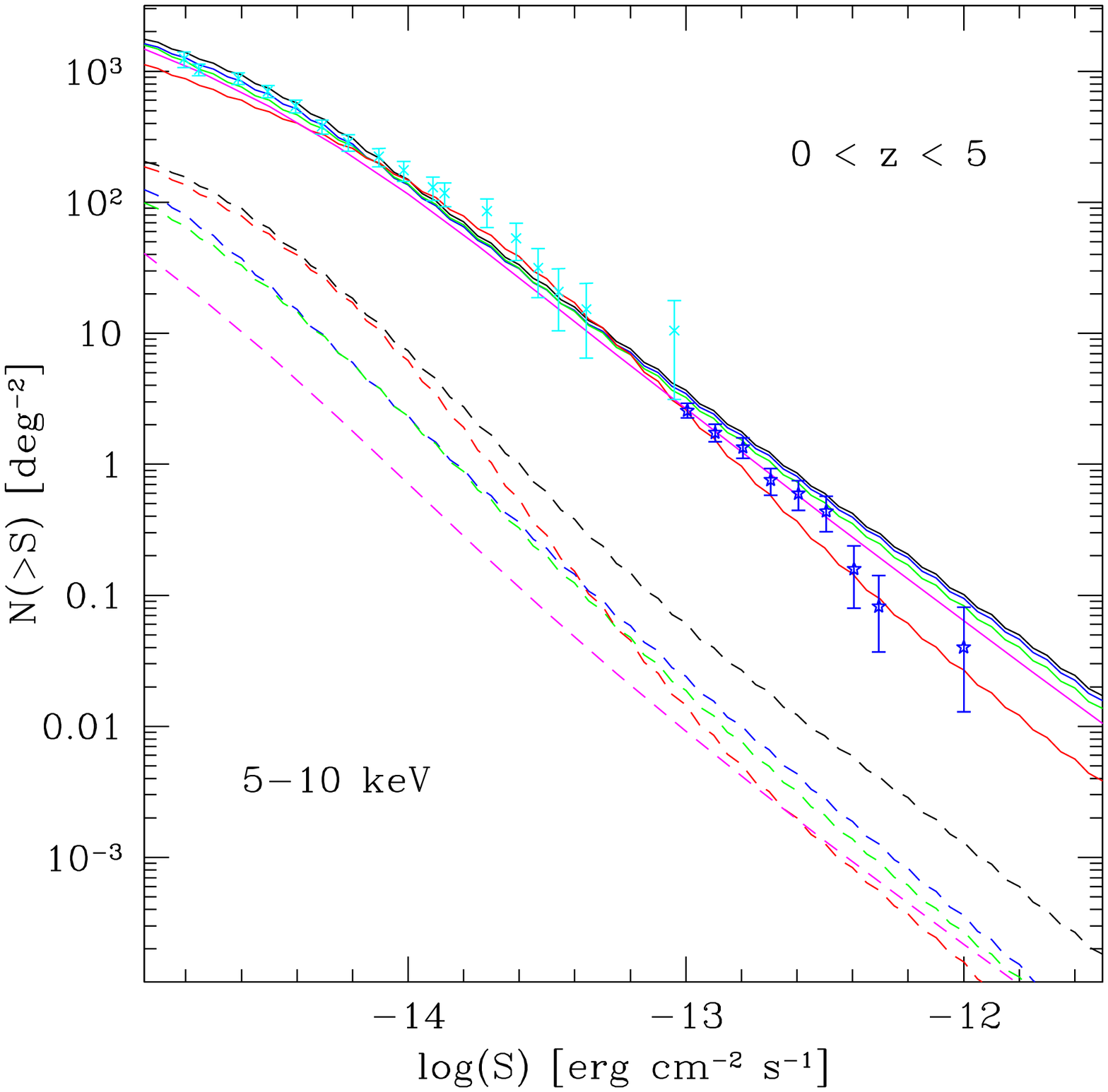}
\caption{Predicted total (solid lines) and Compton thick (dashed
  lines) AGN number counts on the $5$--$10$~keV band. The colors
  denote the different XRB models as described in
  Fig.~\ref{fig:cxrb}. The magenta lines show the predictions of the
  \citet{gch07} model. The cyan crosses are the measured counts from
  the \xmm\ survey of the Lockman Hole \citep{brun08}, and the blue
  stars are from the \xmm\ Hard Bright Serendipitous Survey
  \citep{dc04}.}
\label{fig:5to10kev}
\end{figure}

The hardest energy band that \chandra\ and \xmm\ have measured AGN
number counts is $5$--$10$~\kev\
\citep[e.g.,][]{ros02,dc04,brun08}. For completeness, and to help compare
against the available data, Figure~\ref{fig:5to10kev} plots
predictions of the total and Compton thick AGN $5$--$10$~\kev\ counts for four of the models
discussed in the paper, as well as the \citet{gch07} model (magenta
lines). The Compton thick counts from this last model are bracketed by
the conservative model (not shown) and the \citet{ueda03} model at all
fluxes. The cyan data show the measured counts
obtained by the \xmm\ survey of the Lockman Hole \citep{brun08}, and
the blue points plot the counts from the \xmm\ Hard Bright
Serendipitous Survey \citep{dc04}.   

{}

\clearpage

\clearpage

\clearpage

\clearpage

\clearpage

\clearpage

\clearpage

\clearpage

\clearpage

\clearpage  

\clearpage

\begin{deluxetable}{cccccccc}
\tabletypesize{\footnotesize}
\tablewidth{0pt}
\tablecaption{\label{table:cornershift}\nustar\ Survey Parameters}
\tablecolumns{8}
\tablehead{
\colhead{Depth} & \colhead{Pointings} &
\multicolumn{3}{c}{Avg. Sensitivity (erg cm$^{-2}$ s$^{-1}$)} &
\multicolumn{3}{c}{X-ray Background Fraction} \\ \colhead{(ks)} &
\colhead{\phantom{i}} & \colhead{6--10 keV} & \colhead{10--30 keV} &
\colhead{30--60 keV} & \colhead{6--10 keV} & \colhead{10--30 keV} &
\colhead{30--60 keV}} 
\startdata
\cutinhead{Bo\"{o}tes (9.3 deg$^{2}$)}
\sidehead{Corner Shift}
15.9 & 392 & $2.0\times 10^{-14}$ & $9.9\times 10^{-14}$ & $7.0\times
10^{-13}$ & 0.16 & 0.14 & 0.05 \\
\sidehead{Half Shift}
7.4 & 841 & $2.1\times 10^{-14}$ & $1.1\times 10^{-13}$ & $7.2\times
10^{-13}$ & 0.16 & 0.14 & 0.05 \\
\cutinhead{COSMOS (2 deg$^2$)}
\sidehead{Corner Shift}
86.4 & 72 & $8.6\times 10^{-15}$ & $4.3\times 10^{-14}$ & $3.1\times
10^{-13}$ & 0.28 & 0.23 & 0.07 \\
\sidehead{Half Shift}
36.8 & 169 & $9.4\times 10^{-15}$ & $4.8\times 10^{-14}$ & $3.3\times
10^{-13}$ & 0.28 & 0.20 & 0.07 \\
\cutinhead{AEGIS (0.5 deg$^2$)}
\sidehead{Corner Shift}
346 & 18 & $4.4\times 10^{-15}$ & $2.3\times 10^{-14}$ & $1.6\times
10^{-13}$ & 0.48 & 0.35 & 0.09 \\
\sidehead{Half Shift}
127 & 49 & $5.3\times 10^{-15}$ & $2.7\times 10^{-14}$ & $1.8\times
10^{-13}$ & 0.42 & 0.27 & 0.09 \\
\cutinhead{ECDF-S (0.25 deg$^2$)}
\sidehead{Corner Shift}
778 & 8 & $3.2\times 10^{-15}$ & $1.6\times 10^{-14}$ & $1.2\times
10^{-13}$ & 0.54 & 0.35 & 0.13 \\
\sidehead{Half Shift}
250 & 25 & $4.0\times 10^{-15}$ & $2.0\times 10^{-14}$ & $1.3\times
10^{-13}$ & 0.48 & 0.35 & 0.13 \\
\tablebreak
\cutinhead{GOODS (0.089 deg$^2$)}
\sidehead{Corner Shift}
3110 & 2 & $1.8\times 10^{-15}$ & $9.7\times 10^{-15}$ & $7.2\times
10^{-14}$ & 0.65 & 0.50 & 0.15 \\
\sidehead{Half Shift}
691 & 9 & $2.9\times 10^{-15}$ & $1.4\times 10^{-14}$ & $9.7\times
10^{-14}$ & 0.54 & 0.50 & 0.13 \\
\enddata
\tablecomments{The total exposure time for each survey is 6.2~Ms,
  constituting a 6 months survey (assuming 50\% efficiency). The
  average sensitivity is the flux limit of the survey at 50\% coverage
  (i.e., half the total area). The fraction of the X-ray background at
  that sensitivity is estimated using the \citet{ueda03} model (see
  Sect.~\ref{sub:models}).}
\end{deluxetable}

\clearpage

\begin{deluxetable}{cccccccc}
\tabletypesize{\footnotesize}
\tablewidth{0pt}
\tablecaption{\label{table:halfnum}AGN Detections for Half Shift \nustar\ Surveys}
\tablecolumns{8}
\tablehead{
\colhead{Model} & \colhead{Redshift Range} & \multicolumn{3}{c}{All AGNs} &
\multicolumn{3}{c}{Compton Thick AGNs} \\ \colhead{\phantom{i}} &
\colhead{\phantom{i}} & \colhead{6--10 keV} & \colhead{10--30 keV} &
\colhead{30--60 keV} & \colhead{6--10 keV} & \colhead{10--30 keV} &
\colhead{30--60 keV}} 
\startdata
\cutinhead{Bo\"{o}tes (9.3 deg$^{2}$)}
\citet{ueda03} & $0 < z < 5$ & 284 & 110 & 3 & 6 & 9 & 0 \\
\phantom{i} & $0 < z < 0.5$ & 113 & 65 & 3 & 3 & 7 & 0 \\
\phantom{i} & $0.5 < z < 1$ & 87 & 24 & 0 & 1 & 1 & 0 \\
\phantom{i} & $1 < z < 2$ & 52 & 10 & 0 & 2 & 1 & 0 \\
\phantom{i} & $2 < z < 5$ & 8 & 1 & 0 & 0 & 0 & 0
\\[1em]
\citet{laf05} & $0 < z < 5$ & 280 & 99 & 2 & 5 & 7 & 0 \\
\phantom{i} & $0 < z < 0.5$ & 80 & 47 & 2 & 2 & 5 & 0 \\
\phantom{i} & $0.5 < z < 1$ & 89 & 25 & 0 & 1 & 1 & 0 \\
\phantom{i} & $1 < z < 2$ & 63 & 11 & 0 & 1 & 0 & 0 \\
\phantom{i} & $2 < z < 5$ & 17 & 2 & 0 & 1 & 0 & 0
\\[1em]
\citet{aird10} & $0 < z < 5$ & 338 & 123 & 1 & 10 & 15 & 0 \\
\phantom{i} & $0 < z < 0.5$ & 91 & 56 & 1 & 3 & 10 & 0 \\
\phantom{i} & $0.5 < z < 1$ & 155 & 47 & 0 & 5 & 4 & 0 \\
\phantom{i} & $1 < z < 2$ & 66 & 10 & 0 & 3 & 1 & 0 \\
\phantom{i} & $2 < z < 5$ & 2 & 0 & 0 & 0 & 0 & 0 \\[1em]
DB10 & $0 < z < 5$ & 296 & 126 & 3 & 17 & 19 & 1 \\
\phantom{i} & $0 < z < 0.5$ & 142 & 83 & 3 & 5 & 11 & 1 \\
\phantom{i} & $0.5 < z < 1$ & 92 & 31 & 0 & 7 & 6 & 0 \\
\phantom{i} & $1 < z < 2$ & 54 & 11 & 0 & 4 & 1 & 0 \\
\phantom{i} & $2 < z < 5$ & 8 & 1 & 0 & 1 & 0 & 0
\\[1em]
\citet{tuv09} & $0 < z < 5$ & 250 & 107 & 3 & 1 & 4 & 0 \\
\phantom{i} & $0 < z < 0.5$ & 102 & 66 & 3 & 0 & 3 & 0 \\
\phantom{i} & $0.5 < z < 1$ & 78 & 28 & 0 & 0 & 1 & 0 \\
\phantom{i} & $1 < z < 2$ & 58 & 12 & 0 & 0 & 0 & 0 \\
\phantom{i} & $2 < z < 5$ & 11 & 0 & 0 & 0 & 0 & 0 \\
\tablebreak
\cutinhead{COSMOS (2 deg$^2$)}
\citet{ueda03} & $0 < z < 5$ & 217 & 79 & 2 & 6 & 7 & 0 \\
\phantom{i} & $0 < z < 0.5$ & 59 & 36 & 2 & 2 & 4 & 0 \\
\phantom{i} & $0.5 < z < 1$ & 75 & 24 & 0 & 2 & 2 & 0 \\
\phantom{i} & $1 < z < 2$ & 60 & 12 & 0 & 2 & 1 & 0 \\
\phantom{i} & $2 < z < 5$ & 10 & 1 & 0 & 1 & 0 & 0
\\[1em]
\citet{laf05} & $0 < z < 5$ & 212 & 74 & 2 & 6 & 6 & 0 \\
\phantom{i} & $0 < z < 0.5$ & 43 & 26 & 1 & 1 & 3 & 0 \\
\phantom{i} & $0.5 < z < 1$ & 70 & 24 & 0 & 2 & 2 & 0 \\
\phantom{i} & $1 < z < 2$ & 68 & 14 & 0 & 2 & 1 & 0 \\
\phantom{i} & $2 < z < 5$ & 15 & 3 & 0 & 1 & 0 & 0
\\[1em]
\citet{aird10} & $0 < z < 5$ & 223 & 99 & 1 & 16 & 16 & 0 \\
\phantom{i} & $0 < z < 0.5$ & 44 & 32 & 1 & 3 & 8 & 0 \\
\phantom{i} & $0.5 < z < 1$ & 97 & 45 & 0 & 8 & 7 & 0 \\
\phantom{i} & $1 < z < 2$ & 68 & 16 & 0 & 5 & 2 & 0 \\
\phantom{i} & $2 < z < 5$ & 4 & 0 & 0 & 0 & 0 & 0 \\[1em]
DB10 & $0 < z < 5$ & 230 & 91 & 2 & 19 & 14 & 0 \\
\phantom{i} & $0 < z < 0.5$ & 74 & 46 & 2 & 3 & 6 & 0 \\
\phantom{i} & $0.5 < z < 1$ & 79 & 30 & 0 & 6 & 6 & 0 \\
\phantom{i} & $1 < z < 2$ & 67 & 14 & 0 & 9 & 3 & 0 \\
\phantom{i} & $2 < z < 5$ & 10 & 1 & 0 & 1 & 0 & 0
\\[1em]
\citet{tuv09} & $0 < z < 5$ & 191 & 77 & 2 & 1 & 4 & 0 \\
\phantom{i} & $0 < z < 0.5$ & 51 & 35 & 2 & 0 & 2 & 0 \\
\phantom{i} & $0.5 < z < 1$ & 65 & 26 & 0 & 0 & 1 & 0 \\
\phantom{i} & $1 < z < 2$ & 65 & 14 & 0 & 1 & 0 & 0 \\
\phantom{i} & $2 < z < 5$ & 11 & 1 & 0 & 0 & 0 & 0 \\
\tablebreak
\cutinhead{AEGIS (0.5 deg$^2$)}
\citet{ueda03} & $0 < z < 5$ & 171 & 62 & 2 & 7 & 6 & 0 \\
\phantom{i} & $0 < z < 0.5$ & 34 & 22 & 1 & 1 & 3 & 0 \\
\phantom{i} & $0.5 < z < 1$ & 57 & 22 & 0 & 2 & 2 & 0 \\
\phantom{i} & $1 < z < 2$ & 61 & 13 & 0 & 3 & 1 & 0 \\
\phantom{i} & $2 < z < 5$ & 11 & 2 & 0 & 1 & 0 & 0
\\[1em]
\citet{laf05} & $0 < z < 5$ & 156 & 59 & 1 & 6 & 5 & 0 \\
\phantom{i} & $0 < z < 0.5$ & 26 & 16 & 1 & 1 & 2 & 0 \\
\phantom{i} & $0.5 < z < 1$ & 51 & 20 & 0 & 2 & 2 & 0 \\
\phantom{i} & $1 < z < 2$ & 58 & 15 & 0 & 3 & 1 & 0 \\
\phantom{i} & $2 < z < 5$ & 12 & 3 & 0 & 1 & 0 & 0
\\[1em]
\citet{aird10} & $0 < z < 5$ & 144 & 71 & 1 & 17 & 15 & 0 \\
\phantom{i} & $0 < z < 0.5$ & 24 & 18 & 1 & 3 & 5 & 0 \\
\phantom{i} & $0.5 < z < 1$ & 58 & 32 & 0 & 7 & 7 & 0 \\
\phantom{i} & $1 < z < 2$ & 52 & 17 & 0 & 7 & 3 & 0 \\
\phantom{i} & $2 < z < 5$ & 5 & 0 & 0 & 0 & 0 & 0 \\[1em]
DB10 & $0 < z < 5$ & 184 & 71 & 2 & 19 & 11 & 0 \\
\phantom{i} & $0 < z < 0.5$ & 43 & 28 & 2 & 2 & 3 & 0 \\
\phantom{i} & $0.5 < z < 1$ & 60 & 25 & 0 & 4 & 4 & 0 \\
\phantom{i} & $1 < z < 2$ & 68 & 16 & 0 & 10 & 4 & 0 \\
\phantom{i} & $2 < z < 5$ & 13 & 2 & 0 & 2 & 0 & 0
\\[1em]
\citet{tuv09} & $0 < z < 5$ & 151 & 59 & 2 & 1 & 3 & 0 \\
\phantom{i} & $0 < z < 0.5$ & 29 & 21 & 1 & 0 & 1 & 0 \\
\phantom{i} & $0.5 < z < 1$ & 49 & 22 & 0 & 0 & 1 & 0 \\
\phantom{i} & $1 < z < 2$ & 60 & 15 & 0 & 0 & 1 & 0 \\
\phantom{i} & $2 < z < 5$ & 13 & 1 & 0 & 0 & 0 & 0 \\
\tablebreak
\cutinhead{ECDF-S (0.25 deg$^2$)}
\citet{ueda03} & $0 < z < 5$ & 145 & 54 & 1 & 7 & 5 & 0 \\
\phantom{i} & $0 < z < 0.5$ & 26 & 17 & 1 & 1 & 2 & 0 \\
\phantom{i} & $0.5 < z < 1$ & 48 & 20 & 0 & 2 & 2 & 0 \\
\phantom{i} & $1 < z < 2$ & 54 & 13 & 0 & 3 & 1 & 0 \\
\phantom{i} & $2 < z < 5$ & 12 & 2 & 0 & 1 & 0 & 0
\\[1em]
\citet{laf05} & $0 < z < 5$ & 129 & 51 & 1 & 6 & 5 & 0 \\
\phantom{i} & $0 < z < 0.5$ & 20 & 13 & 1 & 1 & 2 & 0 \\
\phantom{i} & $0.5 < z < 1$ & 42 & 18 & 0 & 2 & 2 & 0 \\
\phantom{i} & $1 < z < 2$ & 49 & 15 & 0 & 3 & 1 & 0 \\
\phantom{i} & $2 < z < 5$ & 11 & 2 & 0 & 1 & 0 & 0
\\[1em]
\citet{aird10} & $0 < z < 5$ & 111 & 58 & 1 & 16 & 13 & 0 \\
\phantom{i} & $0 < z < 0.5$ & 17 & 14 & 1 & 2 & 4 & 0 \\
\phantom{i} & $0.5 < z < 1$ & 43 & 26 & 0 & 6 & 6 & 0 \\
\phantom{i} & $1 < z < 2$ & 43 & 16 & 0 & 7 & 3 & 0 \\
\phantom{i} & $2 < z < 5$ & 5 & 1 & 0 & 1 & 0 & 0 \\[1em]
DB10 & $0 < z < 5$ & 156 & 62 & 2 & 17 & 10 & 0 \\
\phantom{i} & $0 < z < 0.5$ & 32 & 21 & 1 & 1 & 3 & 0 \\
\phantom{i} & $0.5 < z < 1$ & 50 & 22 & 0 & 4 & 3 & 0 \\
\phantom{i} & $1 < z < 2$ & 61 & 17 & 0 & 10 & 4 & 0 \\
\phantom{i} & $2 < z < 5$ & 14 & 2 & 0 & 3 & 0 & 0
\\[1em]
\citet{tuv09} & $0 < z < 5$ & 125 & 52 & 1 & 2 & 3 & 0 \\
\phantom{i} & $0 < z < 0.5$ & 21 & 16 & 1 & 0 & 1 & 0 \\
\phantom{i} & $0.5 < z < 1$ & 40 & 19 & 0 & 0 & 1 & 0 \\
\phantom{i} & $1 < z < 2$ & 51 & 15 & 0 & 1 & 1 & 0 \\
\phantom{i} & $2 < z < 5$ & 13 & 2 & 0 & 0 & 0 & 0 \\
\tablebreak
\cutinhead{GOODS (0.089 deg$^2$)}
\citet{ueda03} & $0 < z < 5$ & 106 & 45 & 1 & 6 & 4 & 0 \\
\phantom{i} & $0 < z < 0.5$ & 17 & 11 & 1 & 1 & 2 & 0 \\
\phantom{i} & $0.5 < z < 1$ & 34 & 16 & 0 & 2 & 2 & 0 \\
\phantom{i} & $1 < z < 2$ & 41 & 14 & 0 & 3 & 1 & 0 \\
\phantom{i} & $2 < z < 5$ & 11 & 2 & 0 & 1 & 0 & 0
\\[1em]
\citet{laf05} & $0 < z < 5$ & 94 & 41 & 1 & 5 & 4 & 0 \\
\phantom{i} & $0 < z < 0.5$ & 13 & 9 & 1 & 1 & 1 & 0 \\
\phantom{i} & $0.5 < z < 1$ & 31 & 14 & 0 & 1 & 1 & 0 \\
\phantom{i} & $1 < z < 2$ & 36 & 13 & 0 & 3 & 1 & 0 \\
\phantom{i} & $2 < z < 5$ & 9 & 2 & 0 & 1 & 0 & 0
\\[1em]
\citet{aird10} & $0 < z < 5$ & 75 & 41 & 1 & 13 & 10 & 0 \\
\phantom{i} & $0 < z < 0.5$ & 10 & 9 & 1 & 2 & 3 & 0 \\
\phantom{i} & $0.5 < z < 1$ & 27 & 17 & 0 & 5 & 5 & 0 \\
\phantom{i} & $1 < z < 2$ & 30 & 13 & 0 & 6 & 3 & 0 \\
\phantom{i} & $2 < z < 5$ & 5 & 1 & 0 & 1 & 0 & 0 \\[1em]
DB10 & $0 < z < 5$ & 115 & 51 & 1 & 15 & 8 & 0 \\
\phantom{i} & $0 < z < 0.5$ & 21 & 14 & 1 & 1 & 2 & 0 \\
\phantom{i} & $0.5 < z < 1$ & 35 & 17 & 0 & 2 & 2 & 0 \\
\phantom{i} & $1 < z < 2$ & 46 & 17 & 0 & 8 & 4 & 0 \\
\phantom{i} & $2 < z < 5$ & 13 & 2 & 0 & 3 & 0 & 0
\\[1em]
\citet{tuv09} & $0 < z < 5$ & 87 & 42 & 1 & 2 & 2 & 0 \\
\phantom{i} & $0 < z < 0.5$ & 13 & 10 & 1 & 0 & 1 & 0 \\
\phantom{i} & $0.5 < z < 1$ & 28 & 15 & 0 & 0 & 1 & 0 \\
\phantom{i} & $1 < z < 2$ & 36 & 14 & 0 & 1 & 1 & 0 \\
\phantom{i} & $2 < z < 5$ & 10 & 2 & 0 & 0 & 0 & 0 \\
\enddata
\tablecomments{These are 4$\sigma$ detections for the half shift
  surveys. Numbers for other tilings/surveys are available from the
  authors. The models are described in Sect.~\ref{sub:models}. The
  three fixed Compton thick fraction models utilize the HXLFs of Ueda \etal\ (2003;
  $f_{\mathrm{CT}}=0.3$), La Franca \etal\ (2005;
  $f_{\mathrm{CT}}=0.4$) and Aird \etal\ (2010;
  $f_{\mathrm{CT}}=0.5$). The DB10 model is the \citet{db10} model of Compton
  thick evolution and makes use of the \citet{ueda03} HXLF. Finally,
  the the most conservative view of the Compton thick
  population is based on the model of \citet{tuv09} and also uses the
  \citet{ueda03} HXLF. Numbers have been
  rounded, so the sum of the individual redshift ranges may not always
equal the value in the $0 < z < 5$ row.}
\end{deluxetable}
\end{document}